\documentclass[apj]{emulateapj}

\shorttitle{Star Formation in Virgo and Isolated Spirals}

\journalinfo{The Astrophysical Journal}
\submitted{Accepted 2004 June 3}
\received{2002 August 23}
\begin{document}

\title{Massive Star Formation Rates and Radial Distributions from 
H$\alpha$ Imaging of 84 Virgo Cluster and Isolated Spiral Galaxies}

\author{Rebecca A. Koopmann}
\affil{Union College}
\affil{Department of Physics, Schenectady, NY 12308}
\email{koopmanr@union.edu}

\author{Jeffrey D. P. Kenney}
\affil{Astronomy Department}
\affil{Yale University, P.O. Box 208101, New Haven, CT 06520-8101}
\email{kenney@astro.yale.edu}

\begin{abstract}

The massive star formation properties of 55 Virgo Cluster and 29
isolated S0-Scd bright ($M_B \leq $ -18) spiral galaxies are compared
via analyses of R and H$\alpha$ surface photometry and integrated
fluxes as functions of Hubble type and central R light concentration
(bulge-to-disk ratio).  In the median, the total normalized massive
star formation rates (NMSFRs) in Virgo Cluster spirals are reduced by
factors up to 2.5 compared to isolated spiral galaxies of the same
type or concentration, with a range from enhanced (up to 2.5 times) to
strongly reduced (up to 10 times).  Within the inner 30\% of the
optical disk, Virgo Cluster and isolated spirals have similar ranges
in NMSFRs, with similar to enhanced (up to 4 times) median NMSFRs for
Virgo galaxies.  NMSFRs in the outer 70\% of the optical disk are
reduced in the median by factors up to 9 for Virgo Cluster spirals,
with more severely reduced star formation at progressively larger disk
radii.  Thus the reduction in total star formation of Virgo Cluster
spirals is caused primarily by spatial truncation of the star-forming
disks.  The correlation between HI deficiency and R light central
concentration is much weaker than the correlation between HI
deficiency and Hubble type. The previously observed systematic
difference in HI spatial distributions and kinematics between early-
and late-type spirals in the Virgo Cluster is at least partially due
to the misleading classification of stripped spirals as early-types.
ICM-ISM stripping of the gas from spiral galaxies is likely
responsible for the truncated star-forming disks of Virgo Cluster
spirals.  This effect may be responsible for a significant part of the
morphology-density relationship, in that a large fraction of Virgo
Cluster galaxies classified as Sa are HI-deficient galaxies with
truncated star forming disks rather than galaxies with large
bulge-to-disk ratios.

\end{abstract}


\section{Introduction}

What role does the environment play in the evolution of cluster galaxies?
The observation that the morphological mix
of galaxies varies in different nearby environments was qualitatively
noted even in the early studies of the Virgo Cluster by Hubble and Humason 
(1931) and has been confirmed in many
studies (e.g., Oemler 1974; Dressler 1980; Postman \& Geller 1984; 
Dressler et al. 1997). In addition, many studies of nearby galaxies
have detailed how 
cluster galaxies differ from field galaxies within the same
Hubble type, including redder colors (Kennicutt 1983a;
Oemler 1992), less HI gas
(Chamaraux, Balkowski, \& G\'erard 1980; Giovanelli \& Haynes 1983), 
and truncated
outer HI gas disks (Giovanelli \& Haynes 1983; Warmels 1988; 
Cayatte et al. 1990). 
Studies of higher redshift cluster galaxies show that evolution in cluster
galaxy morphology and star formation properties has occurred over the
last several billion years. 
Butcher \& Oemler (1978) showed that distant clusters of galaxies have 
a higher proportion of blue galaxies. More recently,
Dressler et al. (1997) found an excess of spirals
and a lack of S0 galaxies in about the same proportion in dense clusters
at redshifts of 0.5 compared to local dense clusters. These results suggest
that many of the S0's in local clusters were actively star-forming spirals
at z=0.5. 

There is a rich literature about the types of environmental processes which
could affect the evolution of galaxies in clusters. There are processes
which affect mainly the gaseous content of a galaxy,
such as ICM-ISM interactions (reviewed by van Gorkom 2004), starvation
(Larson, Tinsley, and Caldwell 1980), 
and gas accretion (Kenney et al., in prep).
Gravitational processes, which affect both the gaseous and stellar properties
of a galaxy, range from low-velocity tidal interactions and mergers, to
high velocity interactions between galaxies and/or the cluster 
(reviewed by Struck 1999 and Mihos 2004). 
Despite a number of recent studies of nearby and distant clusters,
it is not yet clear which processes, if any, are
dominant. Indeed the properties of cluster galaxies may be determined
by a variety of environmental interactions over a Hubble time (Miller
1988; Oemler 1992; Moore et al. 1998). 

The rate of ongoing star formation is
an important measure of the evolutionary state of a galaxy,
and a sensitive indicator of some types of environmental interactions.
Previous studies of the star formation rates of cluster galaxies have 
reached varying and sometimes opposite conclusions.
Some authors have found reduced star formation rates in clusters
(Kennicutt 1983a; Bicay \& Giovanelli 1987; Kodaira et al.
1990; Moss \& Whittle 1993; Abraham et al. 1996; Balogh et al. 1998;
Koopmann \& Kenney 1998; Hashimoto et al. 1998; Gavazzi et al. 2002),
others similar rates (Kennicutt, Bothun, \& Schommer 1984; Donas et al. 1990;
Gavazzi, Boselli, \& Kennicutt et al. 1991, Gavazzi et al. 1998),
and others enhanced rates (Moss \& Whittle 1993, 2000; Bennett \& Moss 1998).
This confused situation on cluster galaxy star formation rates 
is one of the motivations for the present work.
In addition, many previous studies of star formation rates
have been based on aperture or
integrated galaxy spectral observations. Spatial studies of star formation
can probe the types of environmental interactions at work in nearby clusters, 
therefore revealing what may have influenced galaxies in richer clusters
earlier in the history of the Universe.

This work describes results from
an imaging survey of Virgo Cluster and isolated spiral galaxies in both 
broadband R and the H$\alpha$ emission line. 
The intent of this study is to compare
the amounts and distributions of massive star formation in Virgo
Cluster galaxies to those of
a relatively undisturbed isolated sample of galaxies. 
We base our comparisons on 
the H$\alpha$ emission from galaxies, which is a good tracer of the massive
star formation rate (Kennicutt 1983b) in relatively dust-free regions.
It can be used to estimate the total star formation rate by making standard
assumptions for the initial mass function. We use the H$\alpha$ surface
brightness and the H$\alpha$ flux
normalized by the R flux as distance-independent
tracers of the massive star formation rate. 
The data we have gathered thus allow a quantitative radial 
comparison of massive star formation rates in the two environments. 
The Virgo Cluster is a particularly good laboratory to study environmental
effects on star formation since it is the nearest moderately rich cluster,
it has significant ICM, it is dynamically young
with a current infall of spirals 
(Tully \& Shaya 1984), and its galaxies are subject to a variety
of environmental effects.

An important consideration in addressing these issues is the objective
comparison of galaxies with different morphologies. 
van den Bergh (1976) made the important observation,
based on visual inspection of images on plates,
that many cluster spiral galaxies
have low rates of star formation relative to field spirals,
and that the traditional Hubble classification does not work well
in nearby clusters because the bulge-to-disk (B/D) 
ratio is not well correlated
with the disk star formation rate.
Koopmann \& Kenney (1998) confirmed and quantified this effect,
using objective measurements from CCD images,
showing that a one-dimensional classification
scheme such as the Hubble classification is not applied in the same
way to field and cluster galaxies, and that it is not adequate to 
describe the wider range in morphologies of cluster galaxies.
For example, Koopmann \& Kenney find that a
significant fraction ($\sim$ 50\%) of Virgo Cluster spirals classified as Sa
are small-to-intermediate concentration (B/D)
galaxies with reduced global star formation rates, presumably due to
environmental effects. 
This effect contributes to the excess of `early-type' spiral
galaxies in the Virgo
Cluster and therefore to the local morphology-density relationship. 

This paper presents the main results of a program which is
published in separate papers.
We present the observational 
data for the Virgo galaxies in Koopmann et al. (2001, hereafter PI)
and that for the isolated galaxies in Koopmann \& Kenney (2004, in prep, 
hereafter PII).
These papers include H$\alpha$ and R images and radial profiles 
for all galaxies.
Koopmann \& Kenney (1998) 
present comparisons of Hubble type and central R concentration
for isolated and Virgo galaxies. 
The present paper concentrates on the massive star formation
properties, with comparisons between integrated H$\alpha$ fluxes,
H$\alpha$ radial profiles, 
and the relative concentrations of H$\alpha$ emission. 
A comparison of the different types of H$\alpha$ morphologies and a 
discussion of environmental effects is given in Koopmann \& Kenney (2004, 
hereafter PIV).

\section{The Sample}

The 55 Virgo S0-Scd galaxies 
have  $M_B$ $\lesssim$ -18 (B$_T^0$ $\lesssim$ 13 
for an assumed distance of
16 Mpc) and inclinations less than 75$^{\circ}$. The Sa-Scd sample is 95\%
complete for $M_B$ $\leq$ -19 and 68\% complete for $M_B$ $\leq$ -18,
which is about 2 magnitudes fainter than L$^*$. The Sb-Scd sample is
relatively more complete than the Sa-Sab sample (79\% and 50\% to 
$M_B$ $\leq$ -18, respectively). The
S0 galaxies observed are 13\% of the population to $M_B$ $\leq$ -18. Since
this study was not designed to focus on S0's, the results for these galaxies
will not be extensively discussed.
There is a slight bias toward galaxies with higher IRAS fluxes in our
sample, since a subset of these galaxies (75\%)
were obtained as part of a study of star formation rates and efficiencies in
a larger sample of spiral galaxies (Young et al. 1996). 
Observational and reduction details are discussed in PI.

Isolated galaxies were chosen from the Nearby Galaxies Catalog (Tully 1987)
and Gourgoulhon, Chamaraux, \& Fouqu\'{e} (1992) to be 
similar to the Virgo Cluster sample galaxies in B luminosity,
inclination, and distance, but are located in the least dense regions as
defined by parameters from Tully (1987). 
The search criteria produced 103 Sa-Scd galaxies with $M_B \leq $ -18. Our 24
Sa-Scd galaxies comprise $\sim$ 25\% of this sample to $M_B$ $\leq$ -19 and 
$\sim$ 23\% to $M_B$ $\leq$ -18. 
The completeness varies as a function of Hubble type due to the morphology
density relationship, so that S0-S0/a and Sa-Sab galaxies are observed to 
higher completeness ($\sim$ 50\% to $M_B$ $\leq$ -18) than the
Sb-Scd galaxies ($\sim$ 18\% to $M_B$ $\leq$ -18).
Details of the selection of isolated galaxies are given in PII.

All R observations are calibrated to the standard Kron-Cousins
R filter. Images were flux-calibrated using spectrophotometric standard stars
(Massey et al. 1988). 
Extinction coefficients were obtained from Landolt
standards or from standard extinction curves at KPNO and CTIO. 
Continuum-free H$\alpha$ images were derived by subtracting scaled R images
from the line images.
H$\alpha$ in this paper should be read as H$\alpha$+[N II], since two
[N II] lines ($\lambda \lambda$  6584\AA, 6548\AA)
are included in the filter and we make no correction.

\section{Surface Photometry}

Surface photometry of the R and H$\alpha$ images of sample galaxies 
was performed using an IDL-based program (see PI).
R and H$\alpha$ profiles were measured assuming a fixed center, 
inclination, and position
angle, with inclination and position angle derived primarily from the outer 
R isophotes.
Identical elliptical annuli were applied in the H$\alpha$ and
R surface photometry for each galaxy.
Isophotal radii,
isophotal fluxes, and central light concentrations are calculated
from the derived profiles within the IDL program. Exponential R disk
scalelengths were measured from the profiles using the decomposition
routines in the \it fitting \rm programs in IRAF/STSDAS. 
Profiles and fitted parameters for individual galaxies
are presented in PI and PII.

For comparison between galaxies,
all surface brightness profiles were corrected to face on assuming 
complete transparency in the
disk, i.e., by applying the correction 2.5 log \it (a/b) \rm 
to the surface brightness, where \it a/b \rm is the
major-to-minor axis ratio.
All future references to isophotal radii or fluxes integrated within an 
isophotal radius refer to quantities derived from the profiles 
corrected to face-on. 
The assumption of complete
transparency is obviously incorrect, especially at inner radii, but since
extinction values for other galaxies are poorly known, 
no corrections were attempted (see Giovanelli et al. 1994). 
If the extinction properties of Virgo
Cluster and isolated galaxies are similar, 
there should be no systematic error due to inclination
since the Virgo Cluster and 
isolated samples span similar inclination ranges. The most serious errors
in H$\alpha$ integrated fluxes and radial profiles
will occur for highly inclined galaxies. For example,
Young et al. (1996) find that the H$\alpha$ surface brightness significantly
decreases in the mean with
inclination for galaxies with inclinations greater than $\sim$ 70$^{\circ}$.
All galaxies discussed in this paper have an
inclination of 75$^{\circ}$ or less, and 
we find that the measured quantities do not depend on inclination
in this range.

\subsection{Radial Normalization of Profiles}

Outer isophotal radii are direct tracers of the 
size/luminosity of high surface brightness galaxies and 
depend relatively little upon disk 
extinction, since the outer parts of galaxy disks are nearly transparent 
(Giovanelli et al. 1994, 1995).  
Thus, we measured
the outer isophotal radii at 24 R mag arcsec$^{-2}$ in order
to normalize the profiles of galaxies of different sizes and distances.
The radius at the 24 mag arcsec$^{-2}$ isophote in R was selected 
because the signal at this radius dominates the uncertainty in the
sky background for most of the sample galaxies.
(The uncertainty in the sky level 
includes uncertainties due to scattered light, background gradients, and,
in some cases, the effects of small frame size.) 
The values of $r_{24}$ in R are similar to $r_{25}$ in B.

Disk scalelengths were also used to normalize
radial profiles, and comparisons based on the disk scalelength rather than
the isophotal radius produced no significant difference in the results. 
We chose to use the isophotal radius instead of the disk scalelength
primarily because of the difficulty of fitting bulge and disk models
to sample galaxies, particularly in the Virgo Cluster, which
has proportionately more galaxies with complex profiles. 
23\% (13/55) of Virgo
Cluster galaxies have derived disk scalelengths which are uncertain by
$>$ 20\%, compared to 7\% (2/29) of the isolated sample galaxies 
(PI, PII).   
These numbers include 4 galaxies 
for which we were unable to derive disk scalelengths due to
dominant bulge components (the Virgo galaxy NGC 4383 and the isolated NGC 
3414) and/or complex profiles (the Virgo galaxy NGC 4586, which has
a brighter outer disk component and the isolated NGC 2090, which has a lower
surface brightness outer disk component). 
Of the remaining 11 Virgo galaxies with scalelengths more than
20\% uncertain, eight have a higher surface brightness inner disk compared
to the fit to the outer disk (in 3 cases, this appears to
be related to a star formation enhancement) and three have a higher surface
brightness outer disk.

\section{Galaxy Morphology}
\subsection{Hubble Types}
In order to compare properties of the two samples,
we initially binned galaxies as a function of Hubble type. 
Types were extracted from 
Binggeli, Sandage, and Tammann (1985, hereafter BST),  
the Revised Shapley-Ames Catalog
of Bright Galaxies (Sandage \& Tammann 1987, hereafter RSA), 
the Carnegie Atlas (Sandage \& Bedke 1994, hereafter CA), and
the Third Reference Catalog of 
Bright Galaxies (de Vaucouleurs et al. 1991, hereafter RC3).

An examination of the cataloged types indicates problems
in the application of the Hubble system to Virgo Cluster galaxies.
Three Virgo sample galaxies were classified by BST as Sc/Sa or Sc/S0. 
Several Virgo sample galaxies were classified as Sc by BST, but Sa or S0
by RC3. Two galaxies (NGC 4383, NGC 4694) are classified as
amorphous by RSA/BST, but as Sa and S0 by the RC3.
A comparison between the BST/RSA/CA and RC3 shows more
than a half type difference in classification for
12\% of isolated galaxies, but 30\% of Virgo galaxies.
Quantitative comparison between objective tracers of Hubble criteria show
that even some Virgo galaxies which 
have \it consistent \rm classifications in the two catalogs  
do not fit the Hubble system, particularly small B/D systems
consistently classified as Sa and Sab (Koopmann \& Kenney 1998).
See also van den Bergh (1976) and Bothun (1982).

\subsection{R Profiles}
\label{rprofiles}

\begin{figure}[t]
\includegraphics[scale=0.5]{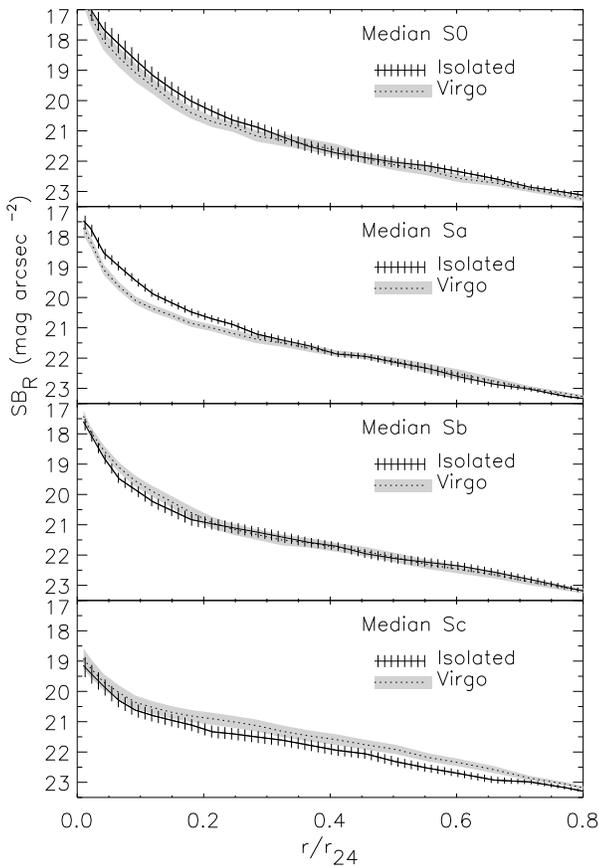}
\caption
{Median R profiles overplotted by type for isolated (solid) and Virgo (dotted)
sample. The range between the 1st and 3rd quartiles is shown by vertical lines
for the isolated sample and shading for the Virgo sample. 
The Virgo Sa median profile is less centrally concentrated than
the isolated Sa median profile. The smaller central concentration of Virgo S0
galaxies is suggestive, but is based on only three Virgo S0 galaxies.
Virgo Sc galaxies tend to have a higher surface brightness in the median.
\label{rmedtype}}
\end{figure}

The R profiles of sample galaxies trace the B/D ratio objectively. 
Median profiles were derived for each Hubble type and environment
bin as a function of $r_{24}$.
A direct comparison between Virgo and 
isolated median profiles for S0-Sc morphological types is
given in Figure~\ref{rmedtype}. 
Interquartile ranges are indicated in the graph with shading.  
Virgo Sa's are less
centrally concentrated in the median than isolated Sa's.
We show in Section~\ref{rconcsect} that this result is significant at the 
99\% level and that it is not caused by a luminosity
effect, as is seen in some studies (e.g., Scodeggio et al. 2002).
This comparison between R radial light profiles further illustrates that Hubble
classifications are not always an indicator of B/D ratio
when applied to the complexity of morphologies in a cluster environment.

The comparison of R profiles also allows an examination of the relative 
surface brightnesses of the sample galaxies.
The extrapolated central surface brightnesses of the
disks of the sample galaxies differ by 
factors of 4-16 for different bins, with no statistically 
significant dependences on either Hubble type or environment. 
This is consistent with previous 
results (e.g., Bothun 1982; Kent 1985) for central disk surface 
brightnesses derived from model fits. 
Figure~\ref{rmedtype} shows that the Virgo Sc galaxies tend
to have higher surface brightness disks than the isolated
galaxies. While Virgo and isolated Sc galaxies span the same
range in surface brightness, there are proportionally more Virgo
Sc's with higher disk surface brightness. 

\subsection{Central R Light Concentration}
\label{rconcsect}
A quantitative indicator of B/D ratio for
galaxies of similar surface brightness is 
the central R light concentration parameter, which can be measured
directly from the R radial profiles and
is ideally independent of the
star formation characteristics of a galaxy.
The central light concentration used in this work is
similar to the parameter used by Abraham et al. (1994):
$$ C30=\frac{F_R(0.3r_{24})}{F_R(r_{24})},$$
where $F_R (r_{24})$ is the total flux in R measured within the $r_{24}$
isophote and $F_R(0.3r_{24})$ is the flux within the 0.3$r_{24}$ isophote.
The $C30$ values are computed directly from the surface photometry
calculated using fixed center, inclination, and position angle. Therefore the
measurement uncertainty is due mainly to uncertainties in the center, 
inclination, and position angle of the elliptical apertures. For
most of the galaxies, these uncertainties are
less than 5\%, which causes a negligible error in $C30$ (PI).
Systematic effects (discussed in detail by Graham et al. 2001) 
are likely more important; for example, any phenomena
(rings, dust, star formation regions) which
change the shape of a galaxy's R surface brightness profile can influence
the measured $C30$ parameter. 

B/D is correlated to $C30$ in a non-linear fashion (Graham 2001). PI provides
a comparison of B/D and $C30$ values.

\begin{figure}[t]
\includegraphics[scale=0.5]{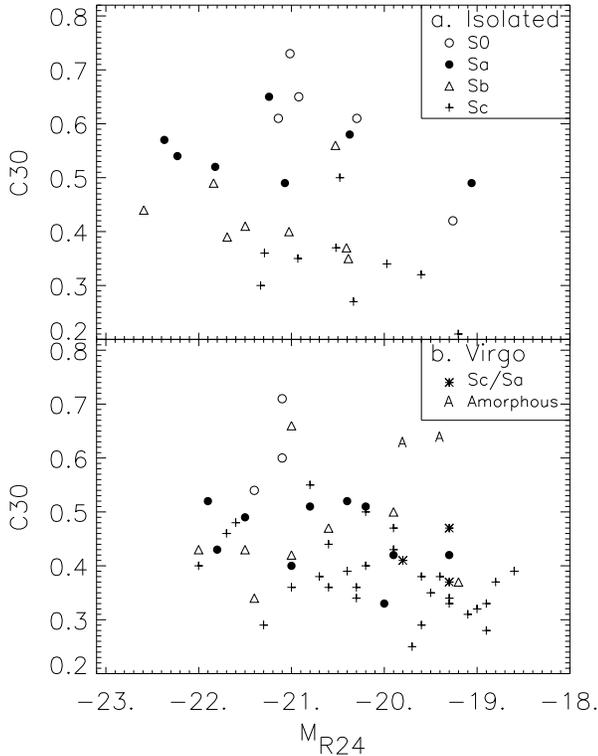}
\caption
{Central R light concentration, $C30$, as a function of luminosity, as 
measured by the absolute R magnitude within the $r_{24}$ isophotal radius
for the isolated (a) and Virgo (b) samples. Symbols
indicate Hubble type. Three galaxies with 
classifications of Sc/Sa or
Sc/S0 are labeled with asterisks. We find no strong correlation between 
the luminosity and $C30$ for galaxies in this sample.
Note that $C30$ and Hubble type are well correlated for the isolated sample,
but poorly correlated for the Virgo sample, showing that Hubble
types do not always trace bulge-to-disk ratios in a cluster environment.
\label{magconc}}
\end{figure}

Scodeggio et al. (2002) find that a near-infrared concentration parameter, 
defined as the ratio between the radii that contain 75\% and 25\% of the 
total light, is strongly dependent on the luminosity of a galaxy, 
particularly for low luminosity galaxies. As shown in  Figure~\ref{magconc},
we find no strong correlations
between $C30$ and the luminosity of a galaxy, as measured by $M_{R24}$, the
R magnitude within the $r_{24}$ isophote. We note that 
our sample is dominated by massive, luminous galaxies, and that we
have a smaller sample of galaxies than Scodeggio et al.

As discussed in Koopmann \& Kenney (1998),
$C30$ and Hubble type are well correlated
in the isolated sample, but not in the Virgo sample, and this can
also be seen in Figure~\ref{magconc}.
In particular, half of the
Virgo galaxies classified as Sa have lower $C30$ than any of the isolated
galaxies classified as Sa. Mann-Whitney tests show that this result is
significant at the 99\% level.

In comparisons between galaxies with different $C30$ values, we bin into
four $C30$ ranges. These ranges were chosen based on comparisons between
Hubble type and $C30$ for the isolated sample, since $C30$ correlates 
well with Hubble type for the isolated galaxies (Koopmann \& Kenney 1998) 
and field galaxies in
general (e.g., Kent 1985). The $C30$ bins correspond roughly to isolated
S0 (0.61 $\le C30 \le$ 0.72), Sa (0.51 $\le C30 \le$ 0.60), Sb (0.38 $\le
C30 \le$ 0.50), 
and Sc (0.24 $\le C30 \le$ 0.37) galaxies. Note that the 
approximately even distribution in Hubble type and 
$C30$ for the isolated galaxies is due to the selection of galaxies, which
was motivated by the need to obtain adequate numbers for statistical
comparison of S0-Sc galaxies, and not to trace the actual 
relative field populations, in which Sc galaxies far outnumber S0-Sa.
In contrast, the Virgo sample galaxies are not evenly distributed in $C30$, 
but primarily fall toward lower concentrations. This is a
combined effect due (i) to a more complete sample of Sc relative
to Sa galaxies compared to the isolated sample, but also (ii) to the fact that
several Virgo galaxies classified as Sa or mixed type have lower 
central concentration than isolated Sa galaxies.

\section{Total Normalized H$\alpha$ Fluxes}
\label{shaflux}

Past comparisons of star formation rates in cluster and field galaxies have
been most often based on integrated measures of star formation. We thus
begin our analysis of the H$\alpha$ images by examining total H$\alpha$ fluxes.

Total H$\alpha$ fluxes were computed from cleaned, sky-subtracted H$\alpha$
images, by summing the total flux within an elliptical aperture 
including all HII regions. 
Uncertainties in integrated H$\alpha$ fluxes due to sky background noise,
calibration errors, continuum subtraction errors are typically 20-30\%. 
The dominant component of this uncertainty is the continuum subtraction error.
Error bars given in this paper include a 3\% 
typical uncertainty in the continuum
subtraction factor, which corresponds to a much larger
error in the calculated H$\alpha$ flux, especially for
galaxies dominated by faint, diffuse emission.
S0 galaxies have the most 
uncertain H$\alpha$ fluxes, since few contain obvious HII regions, 
and we therefore do not place much emphasis on the analysis of their H$\alpha$
properties, although we include them in the plots. 

To compare total H$\alpha$ fluxes between galaxies, we normalized by the flux
in R contained within $r_{24}$ to obtain the normalized massive star formation
rate (hereafter NMSFR or F$_{H\alpha}$/F$_{R24}$).  This quantity 
is a measure of the total H$\alpha$ luminosity per unit red luminosity and
is analogous to an equivalent width. 
Since some readers are more familiar with the equivalent width scale,
we provide a conversion between equivalent width, EW, from our
H$\alpha$ fluxes: 
$$ EW = 1476 (\frac{F_{H\alpha}}{F_{R24}})$$
This equation was derived based on our isolated sample and is further
described in PII. The EW's derived from this correlation
have an uncertainty of 6\% due to scatter.
They correlate well with the
H$\alpha$ equivalent widths measured by Kennicutt and Kent (1983) 
for galaxies in common. 

\begin{figure}[t]
\includegraphics[scale=0.5,angle=90]{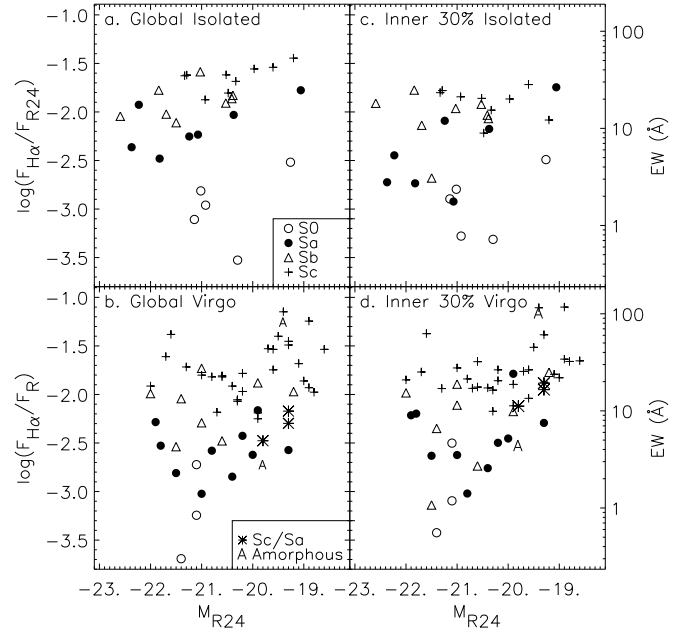}
\caption
{Global and Inner Disk NMSFRs as a function of luminosity, as 
measured by $M_{R24}$, the R magnitude within the $r_{24}$ isophotal radius,
for the isolated (a,c) and Virgo (b,d) samples. 
The right y-axis provides the equivalent width scale.
There is a weak correlation for isolated galaxies, but in Virgo, the 
scatter is greater due to reductions in star formation. A number of Virgo 
spirals have enhanced total and inner NMSFRs compared to the isolated.
\label{maghaflux}}
\end{figure}

As a function of Hubble type 
or $C30$ (see Figure~\ref{maghaflux}a and b), 
galaxies in both samples show a large
spread in total NMSFRs. Within a Hubble type, 
F$_{H\alpha}$/F$_{R24}$ ranges by a factor
of 10, which is similar to measurements in previous surveys of galaxy
star formation rates as a function of Hubble type (Kennicutt 1998 and 
references therein). Within concentration bins, however, the 
F$_{H\alpha}$/F$_{R24}$ of Virgo Cluster galaxies ranges by
factors up to 20-40 (see Section~\ref{integ}). While the distributions 
are non-Gaussian and span a large range, the larger spread in star formation
rates in Virgo is mostly due to galaxies with reduced star formation, 
but is
also partially due to \it enhancements \rm in the NMSFRs by
factors of up to 3 for several galaxies.

Most of the Virgo galaxies with enhanced NMSFRs are lower luminosity
galaxies ($M_B$ $<$ -18) and we do not have a large number of isolated 
spirals of similar luminosity. 
Several authors (e.g., Kennicutt et al. 1984; Boselli et al. 2001) have
shown a tendency for lower luminosity galaxies to exhibit higher H$\alpha$
equivalent widths and star formation rates. A weak correlation is evident
in Figure~\ref{maghaflux}, particularly for the isolated galaxies.
In the Virgo Cluster, the correlation is weaker due mainly to reductions in
star formation. 
To avoid a possible luminosity bias in our analysis, 
we will limit most of our discussions to
galaxies with $M_{R24}$ $\le$ -19.5 ($M_B$ $\le$ -18.5).

Median total normalized H$\alpha$ fluxes with standard deviations and
corresponding EW's for Virgo Cluster and
isolated galaxies with $M_{R24} \le $ -19.5 
are listed in Table~\ref{tab1flux} as a function
of type and central light concentration. In both
samples, the typical NMSFRs decrease as a 
function of earlier Hubble type and of greater central concentration.

For the isolated galaxies,
the median NMSFRs for spiral galaxies of different Hubble
types and concentrations are similar within a factor of 4, 
consistent with other studies of globally averaged
H$\alpha$ surface brightness for larger samples of field galaxies (Young
et al. 1996). 
The tendency for the early-type/high concentration 
(0.51 $ \le C30 \le$ 0.60; approximately Sa) isolated spirals
to have lower NMSFRs compared to the lowest
concentration galaxies is significant ($>$ 99\%) when the binned
samples are compared via Mann-Whitney tests. 

\begin{deluxetable*}{llclcc}
\tabletypesize{\scriptsize}
\tablecaption{Median Normalized Massive Star Formation Rates}
\tablewidth{0pt}
\tablehead{
\colhead{} & 
\colhead{Isolated (\#)}&
\colhead{Isolated}&
\colhead{Virgo(\#)}& 
\colhead{Virgo}& 
\colhead{} \\
\colhead{Type/$C30$}& 
\colhead{log $\frac{F_{H\alpha}}{F_{R24}}$} & 
\colhead{H$\alpha$ EW} & 
\colhead{log $\frac{F_{H\alpha}}{F_{R24}}$} & 
\colhead{H$\alpha$ EW} & 
\colhead{$ \rm \frac{Isolated}{Virgo} $}\\
\colhead{}& 
\colhead{(NMSFR)} & 
\colhead{(\AA)} & 
\colhead{(NMSFR)} & 
\colhead{(\AA)} & 
\colhead{}}
\startdata\\
\multicolumn{6}{c}{\bf Total \rm}\\
\cline{1-6}
S0-S0/a&-2.96 $\pm$ 0.29 (5)&2 &-3.24 $\pm$ 0.49 (3)&$<$ 1&-\\
Sa-Sab& -2.24 $\pm$ 0.20 (6) &9& -2.57 $\pm$ 0.24 (9)& 4 &2.1 (0.8)*\\
Sb-Sbc&-1.87 $\pm$ 0.15 (8) &20&-2.06 $\pm$ 0.24 (7) & 13 &1.5 (0.5) \\
Sc-Scd&-1.63 $\pm$ 0.09 (7) &35& -1.81 $\pm$ 0.19 (18) & 23 &1.5 (0.2)\\
Sc-Scd $<$ 6$^{\circ}$&\ \ \ \ \ \ \ \ -&-&-1.82 $\pm$ 0.28 (14)&22& 1.5 (0.2)\\

.61 $\le C30 <$ 0.73& -2.96 $\pm$ 0.48 (5) &2& -2.76 $\pm$ 0.38 (3)&3& -\\
.51 $\le C30 \le$ 0.60&-2.01 $\pm$ 0.16 (5) &14&-2.41 $\pm$ 0.40 (5)&6& 2.5 (1.7)\\
.38 $\le C30 \le$ 0.50&-2.03 $\pm$ 0.25 (8) & 14 &-1.99 $\pm$ 0.29 (21)& 15 & 0.9 (0.3)\\
.24 $\le C30 \le$ 0.37& -1.65 $\pm$ 0.10 (8) &33&-1.80 $\pm$ 0.18 (9)&23& 1.4 (0.3)\\
\cline{1-6}
\multicolumn{6}{c}{\bf Inner 30\% \rm}\\
\cline{1-6}
Sa-Sab & -2.55 $\pm$ 0.32     & 4& -2.48 $\pm$ 0.41 & 5&0.9 (0.6)\\
Sb-Sbc      & -2.00 $\pm$ 0.14     & 15 &-2.12 $\pm$ 0.20  & 11& 1.3 (0.6)\\
Sc-Scd      & -1.86 $\pm$ 0.10     & 20&-1.86 $\pm$ 0.16 & 20 &1.0 (0.2)\\
Sc-Scd $<$ 6$^{\circ}$ & \ \ \ \ \ \ \ \ - & -& -1.83 $\pm$ 0.28 & 22 & 0.9 (0.2)  \\
.61 $\le C30 <$ 0.73& -2.89 $\pm$ 0.54  &2&-2.57 $\pm$ 0.38 &4 & -\\
.51 $\le C30 \le$ 0.60  & -2.44 $\pm$ 0.37  &5 & -2.48 $\pm$ 0.44  & 5 &1.1 (1.0)\\
.38 $\le C30 \le$ 0.50  & -2.14 $\pm$ 0.23  &11  & -1.91 $\pm$ 0.21 & 18 &0.6 (0.2)\\
.24 $\le C30 \le$ 0.37 & -1.87 $\pm$ 0.09  &20   & -1.94 $\pm$ 0.16 & 17&1.2 (0.2) \\
\cline{1-6}
\multicolumn{6}{c}{\bf Outer 70\%\rm}\\
\cline{1-6}
Sa-Sab & -2.03 $\pm$ 0.25     & 14&-2.98 $\pm$ 0.48  & 2 &9 (7)*\\      
Sb-Sbc      & -1.84 $\pm$ 0.20  & 21&  -2.00 $\pm$ 0.31 & 15&1.5 (0.6)\\
Sc-Scd      & -1.56 $\pm$ 0.09  & 41  & -1.82 $\pm$ 0.24 & 22 & 1.8 (0.3)* \\
Sc-Scd $<$ 6$^{\circ}$ & \ \ \ \ \ \ \ \ -  &-   & -1.83 $\pm$ 0.28 & 22&1.9 (0.2)\\
.61 $\le C30 <$ 0.73& -3.04 $\pm$ 0.31 &1 & -4.00 $\pm$ 2.2 &$<$ 1& -\\
.51 $\le C30 \le$ 0.60   & -1.89 $\pm$ 0.18 & 19    & -2.34 $\pm$ 0.36  & 7&2.8 (1.8)\\
.38 $\le C30 \le$ 0.50 & -1.93 $\pm$ 0.29  &17   & -2.12 $\pm$ 0.44  & 7&1.5 (0.8)*\\
.24 $\le C30 \le$ 0.37 & -1.58 $\pm$ 0.11 &39    & -1.83 $\pm$ 0.21 & 22& 1.8 (0.4) \\
\enddata
\label{tab1flux}
\tablecomments{Median and standard deviation (from the median) of log 
$\frac{F_{Ha}}{F_{R24}}$ and corresponding equivalent width,
as a function of environment, Hubble type, and central R light concentration 
for the isolated and Virgo galaxies with $M_{R24} \leq $-19.5. 
The number of galaxies in each
bin is given in parenthesis after the total flux. Note the larger spread in the
Virgo distribution relative to the isolated distribution.
The ratio of isolated to Virgo rates is given in the 6th column, with the
standard deviation in the median given in parantheses. A number
greater than 1 indicates a reduction for Virgo galaxies, 
while a number less than 1 indicates
an enhancement. An asterisk indicates that the difference in the distributions
is 
significant at better than 99\% according to a Mann-Whitney ranked-sum 
test. 
Virgo Cluster galaxies have reduced total and outer disk star formation
rates in the median for all Hubble type and $C30$ bins. However, the inner
disk rates for Virgo Cluster galaxies are similar to mildly enhanced 
compared to isolated counterparts.}
\end{deluxetable*}

In the median, Virgo 
galaxies have reduced star formation with respect to isolated galaxies
in every Hubble and $C30$ bin, with the exception of the intermediate
$C30$ bin. The larger spread in the Virgo NMSFRs is evident in the 
larger standard deviations. 
Table~\ref{tab1flux} also provides the factor by which Virgo
normalized star formation is reduced in the median with respect to the 
isolated. The error associated with the scatter in normalized star
formation rates is given in parentheses. 
The reduction in star formation as a function of Hubble type is
in agreement with a number of past studies of star
formation rates in the Virgo Cluster 
(e.g., Kennicutt 1983a; Gavazzi et al. 2002).

\section{Radial Distributions of Massive Star Formation}

\label{harad}

Our study of total NMSFRs shows that Virgo Cluster galaxies typically
have less total star formation than isolated counterparts, but that there
is a large range in total star formation rates among cluster galaxies,
which includes both reductions and enhancements.
Using the spatial distributions of H$\alpha$ emission, 
it can be determined 
where in the disks the star formation has been enhanced or reduced.

\subsection{H$\alpha$ Radial Profiles}

\begin{figure}[t]
\includegraphics[scale=0.45]{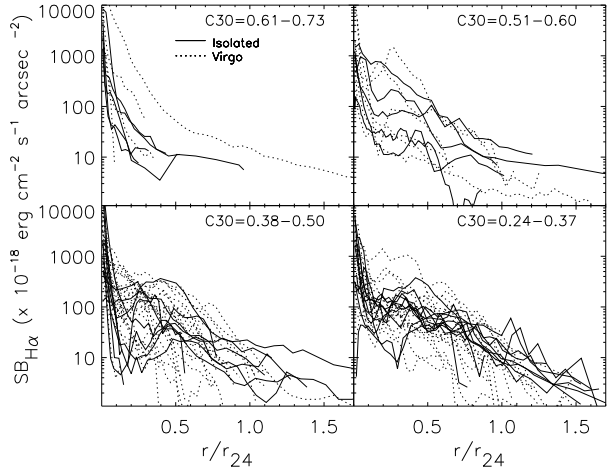}
\caption{
H$\alpha$ profiles, normalized by $r_{24}$, binned by $C30$ of the indicated
ranges and overplotted
for the two environments. Virgo
galaxies, represented by dotted lines, show truncated H$\alpha$ disks compared
to isolated galaxies of similar R light concentration. Several galaxies
have emission in the inner 0.5$r_{24}$ which is
enhanced compared to isolated counterparts.
 \label{haoverconcr24}}
\end{figure}

The H$\alpha$ surface brightnesses of sample galaxies 
in each Hubble class or concentration bin (Figure~\ref{haoverconcr24}), 
span a factor 10 to 100 range at different radii in the disk.
This large range in H$\alpha$ surface brightness is in contrast to the R
surface brightness, which spans a factor 4 to 16 range at any given radius
(Section~\ref{rprofiles}) and to global star
formation rates which vary by about a factor of
10 within a given Hubble type (Section~\ref{shaflux}; Kennicutt 1998). 
The much larger range in H$\alpha$ surface brightness is 
probably associated with variations
in the surface density and critical density of gas as a function
of radius in galaxies (Kennicutt 1989). 
The H$\alpha$ surface brightness distribution 
within an individual galaxy can vary due to a variety of effects,
including bars, rings, spiral arms, 
circumnuclear regions, and the truncation of the
star-forming disk.

A closer look at Figure~\ref{haoverconcr24} reveals several clues
about the relative spatial distributions of star formation in isolated
and Virgo galaxies, despite the large range in H$\alpha$ surface brightnesses.
Many Virgo Cluster galaxies have 
truncated H$\alpha$ disks compared to isolated counterparts.
The smaller H$\alpha$ disks of Virgo Cluster galaxies are 
seen as the emergence of dotted lines, many steeply declining, below the
solid lines beyond about 0.4$r_{24}$ in the plot. 
Truncated profiles are seen in all Hubble type and $C30$ bins.
52\% (27/52) of Virgo 
Cluster spirals have H$\alpha$ disks truncated within 0.8$r_{24}$, compared to
12\% (3/24) of isolated spirals (see PIV). In contrast,
within the truncation radii, the Virgo Cluster galaxies have H$\alpha$
surface brightnesses which are typically similar or enhanced compared to
isolated counterparts. Few Virgo Cluster galaxies have star
formation which is reduced across the disk below isolated rates. 

\begin{figure*}[t]
\hbox{
\includegraphics[scale=0.5]{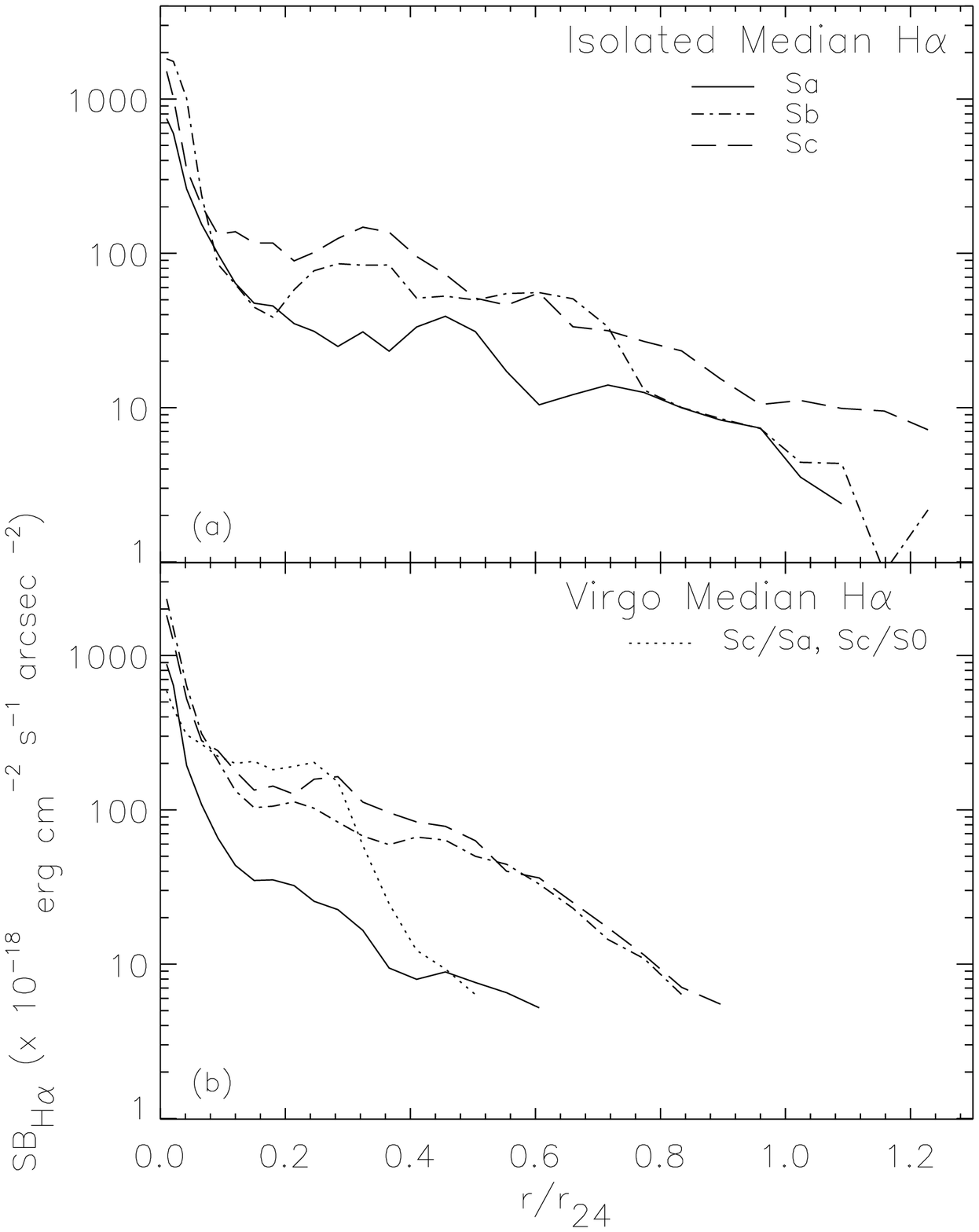}
\includegraphics[scale=0.5]{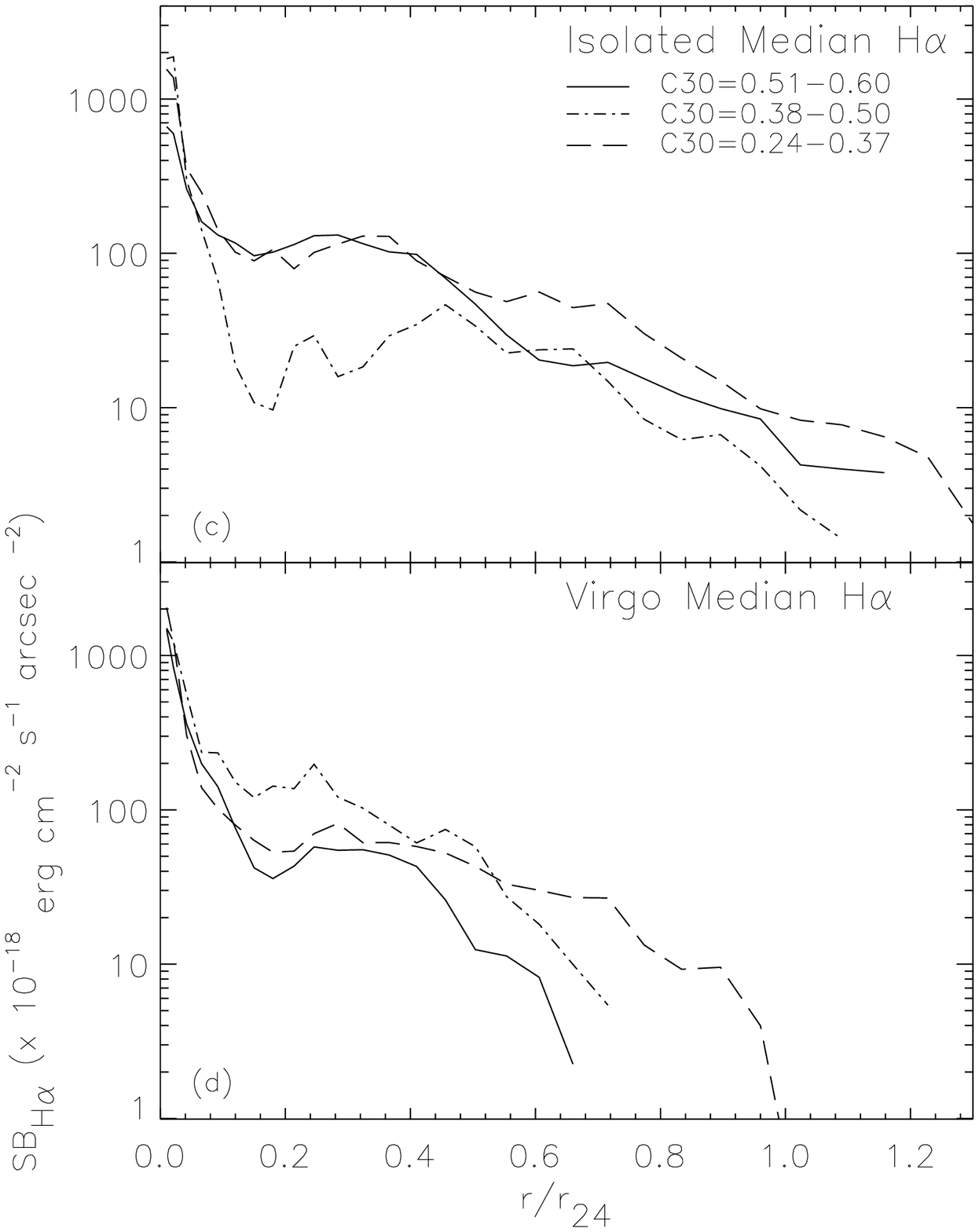}}
\caption{Median H$\alpha$ radial profiles for isolated  and Virgo 
galaxies with $M_{R24} \leq $  -19.5
as a function of Hubble type (a,b) and $C30$ (c,d). 
Despite the scatter by factors of 10-100 in the individual profiles,
isolated median profiles for different Hubble types or different
$C30$ agree to within a factor of 3 over most of the galaxy disk. 
Virgo Cluster median profiles show the effect of
truncation in the H$\alpha$ disks.\label{hamed}}
\end{figure*}

\begin{figure*}
\hbox{
\includegraphics[scale=0.5]{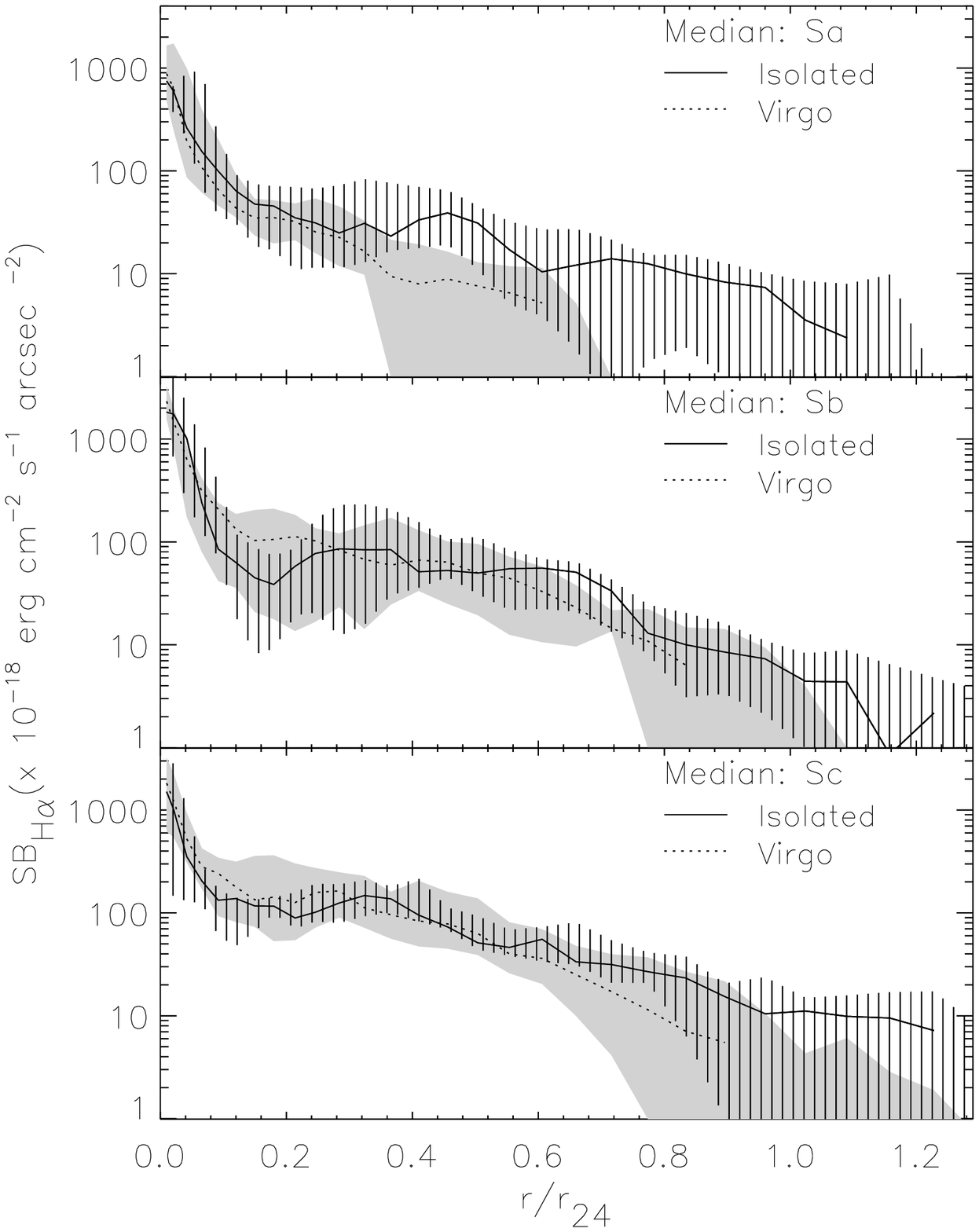}
\includegraphics[scale=0.5]{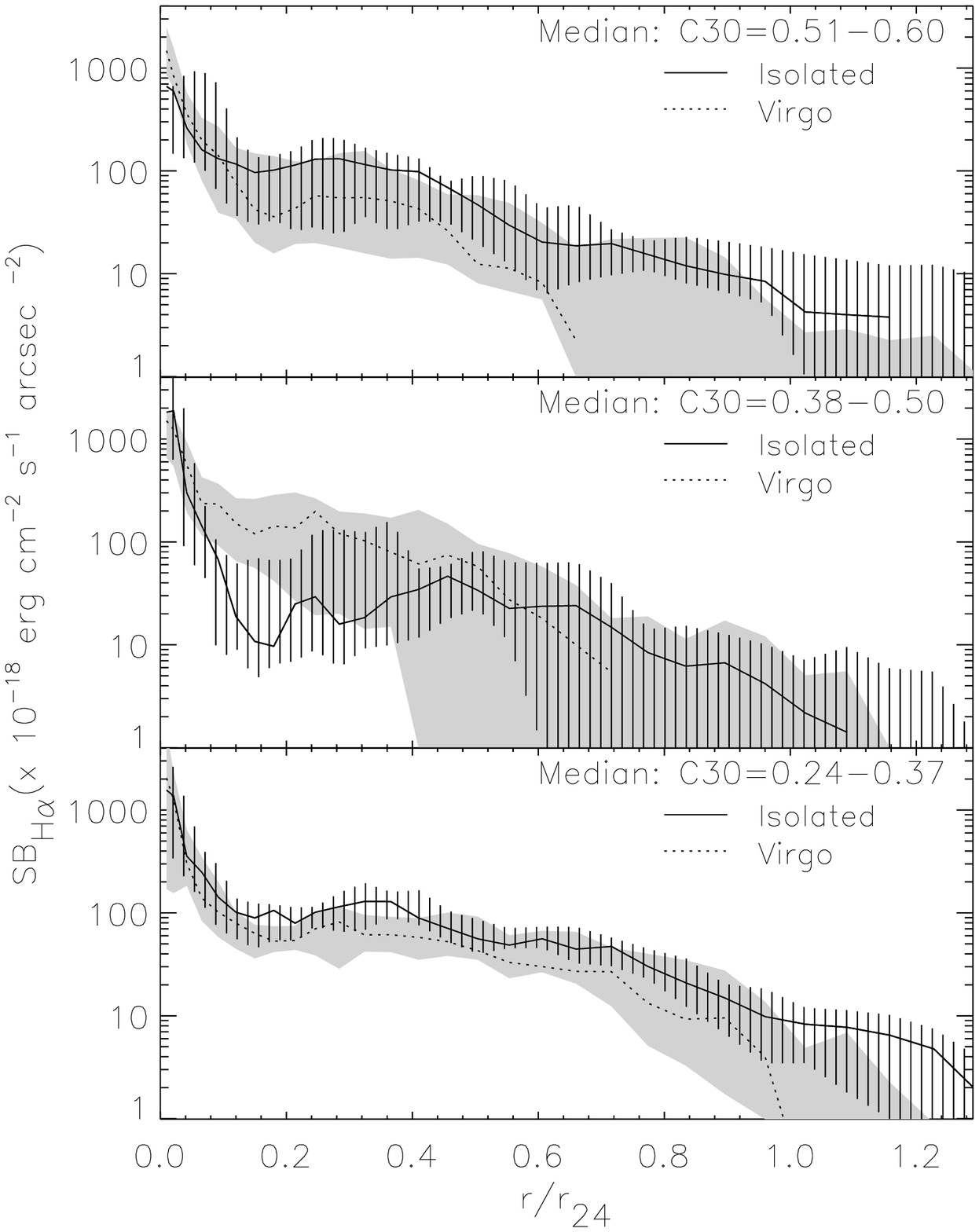}}
\caption{Median H$\alpha$ radial profiles for isolated (solid) and Virgo 
(dotted) galaxies with $M_{R24} \leq $ -19.5
as a function of Hubble type (left) and $C30$ (right). 
The interquartile ranges are shown as vertical lines
for the isolated galaxies and as shading for the Virgo Cluster galaxies.
The Virgo Cluster galaxies have truncated star-forming disks in all bins.
The inner regions of Virgo spirals have normal to enhanced NMSFRs
in most bins, compared to isolated profiles.\label{hamedtype}}
\end{figure*}

To compare profiles in the two samples in a more statistical way, median
profiles were derived for each Hubble type and $C30$ bin, by 
calculating the 
median of the surface brightness at numerous radii across the disk 
for galaxies within a given $C30$ or 
Hubble type bin and with $M_{R24} \leq $ -19.5. 
Outside of the radius of the outermost HII region, the surface brightness
was assumed
to drop to 0 and was counted in subsequent radial bins as 0. Hence, the
median profiles include the effect of truncated H$\alpha$ surface brightness
profiles. Median radial
profiles of different Hubble types and $C30$ are shown for each environment in
Figures~\ref{hamed} and~\ref{hamedtype}. Figure~\ref{hamedtype} includes
shading indicating the interquartile range of the 
surface brightness distributions. The median profiles are drawn for radii
for which the surface brightnesses are greater than 0, but the shading
extends beyond these radii to show the upper ranges in the distributions.

Despite the range of 1-2 orders of magnitude in H$\alpha$ surface brightnesses
at a given radius,
the medians of isolated spirals as a function of Hubble type and $C30$ 
agree to within a factor of 4 over most
of the disk, as shown in Figure~\ref{hamed}. 
The agreement of median radial profiles within a factor of 4
is similar to the agreement between the total median 
H$\alpha$ fluxes (Section~\ref{shaflux}; Young
et al. 1996) and emphasizes that even early-type/high $C30$ isolated
galaxies typically have substantial amounts of ongoing massive star formation
across the disk (see also Hameed \& Devereux 1999).

The most striking difference between the Virgo and isolated median profiles
are the smaller H$\alpha$ disks of Virgo Cluster spirals, which
can be seen in all bins in Figure~\ref{hamedtype}. 
The Virgo H$\alpha$ median profiles are truncated at progressively 
smaller disk radii
for progressively earlier Hubble types and higher $C30$ values. However, there
is a large variation in truncation radii for Virgo galaxies in all Hubble type
and $C30$ bins, as can be seen from the interquartile ranges.
In contrast, within the radii where truncation
becomes important,
the medians of Virgo profiles 
are within a factor of 3 of each other and isolated median profiles for all
$C30$ and Hubble type bins.

There is a tendency for some of the median H$\alpha$ profiles, particularly
the isolated, to have a `dip' in the inner disk surface brightness.
The dip is caused by the presence of barred galaxies in the samples. The
effect is stronger in the isolated bins (note particularly the 0.38-0.50 $C30$ 
isolated bin) because
our isolated sample has a higher occurrence of strong bars than our Virgo
sample (based on
either RSA or RC3 bar classifications). This is not representative of 
the complete
sample from which the isolated and Virgo Cluster galaxies were drawn,
in which strong bars occur at similar frequency. To 
check the differences between star formation characteristics
in the inner disks of barred and unbarred galaxies, 
we compared median
profiles and overplotted individual H$\alpha$
profiles of barred and unbarred galaxies of different types and
$C30$ (not shown). We find in the median that barred galaxies do have lower star
formation rates within 0.2-0.3 $r_{24}$, but that these
lower rates are often compensated by higher rates within 0.3-0.6 $r_{24}$. 

\subsection{Integrated H$\alpha$ Fluxes}
\label{integ}

\begin{figure*}[t]
\includegraphics[scale=0.65,angle=90]{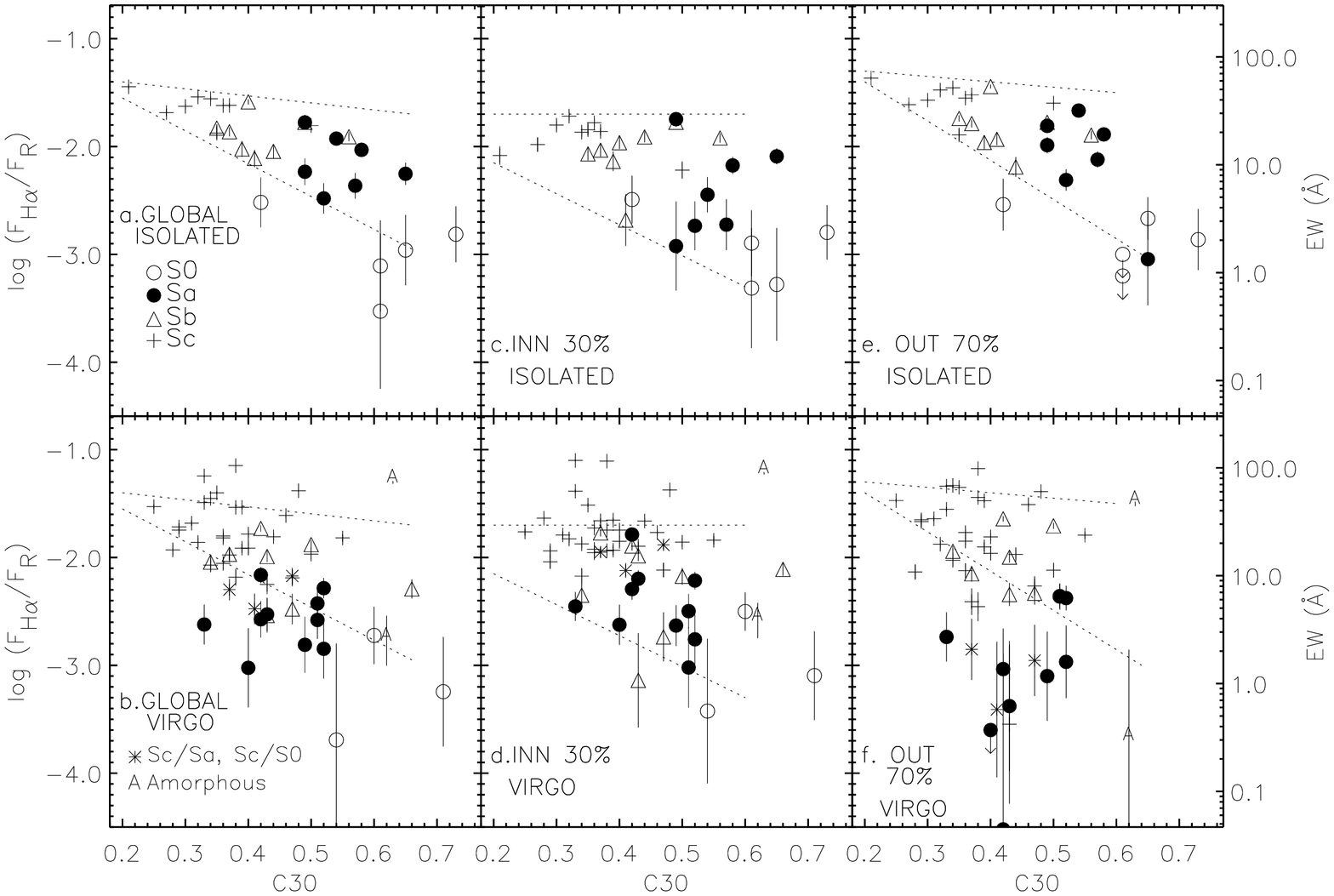}
\caption{Normalized NMSFRs as a function of $C30$. The values for the total
star formation disk are given in a (isolated) and b (Virgo), the values for 
the inner 30\% of the optical disk in c (isolated) and d (Virgo), and the
values for the outer 70\% of the disk in e (isolated) and f (Virgo).
The right y-axis provides the equivalent width scale.
Error bars are derived from the random sky error combined in quadrature
with an estimate of the continuum subtraction error. Where no error bars
appear, the error bar is smaller than the symbol size.
The lines indicate the 
approximate bounds of isolated Sa-Sc galaxies. Typically, Virgo spirals with
reduced star formation have reduced outer star formation, but similar or 
enhanced inner
star formation compared to isolated galaxies. Thus the main cause of
the reduction of star formation is truncation of the star-forming disk.
\label{6pan}}
\end{figure*}

\begin{figure*}
\includegraphics[scale=0.45,angle=90]{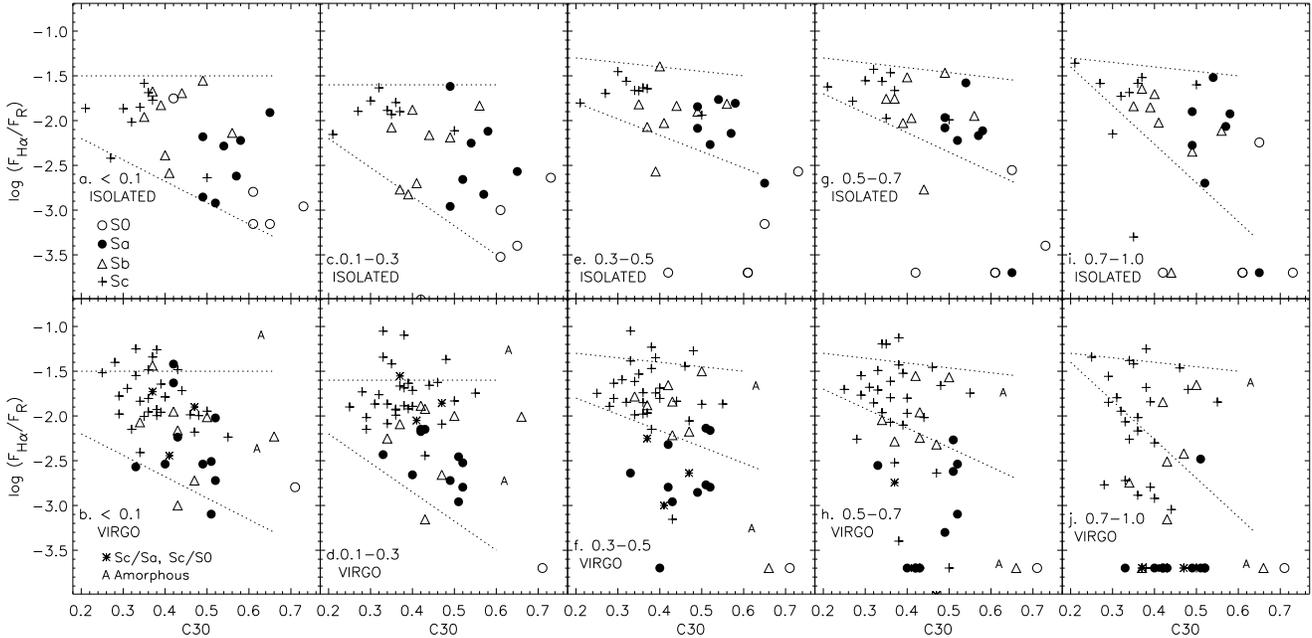}
\caption{Normalized NMSFRs as a function of $C30$ for the indicated bins. 
Symbols which lie at the base of the plot represent values which are 
below the lower y limit of the plot. The lines indicate the 
approximate bounds of isolated Sa-Sc galaxies.
Within 10\% of the optical disk and between 10 and 30\% of the optical
disk, Virgo galaxies have normal to enhanced normalized H$\alpha$ fluxes.
Between 30 and 50\% of the optical disk, 15 of the Virgo galaxies have
reduced star formation. Note that most of the Virgo galaxies classified
as Sa have reduced star formation in this radial bin.
The numbers of Virgo galaxies with star formation rates less than 
their isolated
counterparts progressively increases as larger radius radial bins are 
examined. Even many of the Sc galaxies have truncated outer disks. Notice
the lack of earlier-type galaxies with similar NMSFRs to
isolated in the outer bins.\label{6pancont}}
\end{figure*}

The radial dependence of massive NMSFRs can also be studied
by deriving integrated H$\alpha$ fluxes normalized by R over several radial
bins. Figure~\ref{6pan}
shows the integrated NMSFRs
as a function of environment and $C30$ over the whole disk (a,b), 
the
inner 30\% of $r_{24}$ (c,d), and the outer 70\% of $r_{24}$ (e,f).
Figure~\ref{6pancont}
shows the integrated rates for five smaller 
radial bins: within 0.1$r_{24}$, 0.1$r_{24} <$ r $<$ 0.3$r_{24}$, 
0.3$r_{24}$ $< r <$ 0.5$r_{24}$, 0.5$r_{24} <$ r $<$ 0.7$r_{24}$,
and $<$ 0.7$r_{24} < r <$ 1.0$r_{24}$.
In both figures,
the dotted lines indicate the approximate bounds of isolated Sa-Sc galaxies
for each bin.

Median NMSFRs were calculated for galaxies with $M_{R24} \leq $ -19.5
in each radial bin and are
presented in Tables~\ref{tab1flux} and~\ref{tab2flux}.
Table~\ref{tab1flux} gives the median, standard
deviations (based on the median), and the isolated-to-Virgo ratio for
the $r <$ 0.3 $r_{24}$ and 0.3$r_{24} < r <$ 1.0 $r_{24}$ outer bins.
A comparison of the standard deviations shows that the inner Virgo NMSFRs
have similar values compared to the isolated sample, except in the 0.38-0.50
$C30$ bin, in which the median is enhanced by a factor of 1.7.
In the outer disk, however, the Virgo NMSFRs are reduced
by factors of 1.5 - 9 for galaxies of similar $C30$ or Hubble type. Note
especially the values of 2.8 - 9 for early Hubble type or high $C30$. 
The difference
between the outer disk Sa and 0.51 $\leq C30 \leq$ 0.60 bins is due to
(i) the shifting of several galaxies classified as Sa to lower $C30$ bins
and (ii) the presence of a few galaxies classified as Sc with active star 
formation and higher $C30$.

\begin{deluxetable*}{lccccc}
\tabletypesize{\scriptsize}
\tablecaption{Virgo Star Formation Rates Relative to Isolated}
\tablewidth{0pt}
\tablehead{
\colhead{Type or $C30$}& 
\multicolumn{5}{c}{$ \rm \frac{Isolated}{Virgo}$ Flux Ratio in each radial bin}\\ 
\colhead{}& 
\colhead{$ r <$ 0.1$r_{24}$} & 
\colhead{0.1$r_{24} < r <$ 0.3$r_{24}$} & 
\colhead{0.3$r_{24} < r <$ 0.5$r_{24}$} & 
\colhead{0.5$r_{24} < r <$ 0.7$r_{24}$} &
\colhead{0.7r $_{24} < r <$ 1.0$r_{24}$}}
\startdata
Sa-Sab&0.9 (1.4) & 0.8 (0.4) &5.0 (3.0) & $>$ 5  & $>$ 5 \\
Sb-Sbc & 1.8 (0.6) & 0.7 (0.2) &0.9 (0.4) & 1.6 (0.9) & $>$ 4\\
Sc-Scd & 1.3 (0.4)& 1.0 (0.2) &1.4 (0.3)  & 1.4 (0.3) & 4 (2)\\
.51 $\le $C30 $\le$ 0.60  & 0.8 (0.6) & 1.4 (1.8) & 1.9 (0.7) & $>$ 3  & $>$ 3 \\
.38 $\le C30 \le$ 0.50  & 0.9 (0.4)& 0.4 (0.1)  &0.8 (0.4)  & 2 (1) & $>$ 5\\
.24 $\le C30 \le$ 0.37 & 1.4 (0.5) & 1.3 (0.2) &1.4 (0.2) & 1.4 (0.2) & $>$ 3 \\
\enddata
\label{tab2flux}
\tablecomments{The ratio of isolated to Virgo H$\alpha$ median fluxes in 
the indicated radial bins 
as a function of Hubble type, and central R light concentration for galaxies
with $M_{R24} <$ -19.5. 
In parentheses is given the standard deviation in the median. Lower limits
were calculated based on the background sky error in each image.
Virgo Cluster galaxies have larger median reductions in 
NMSFRs at progressively larger radii
for all Hubble type and $C30$ bins. In the inner disk, ratios show that
Virgo and isolated spirals have similar or enhanced NMSFRs. 
}
\end{deluxetable*}

Table~\ref{tab2flux} gives the median isolated-to-Virgo ratios and standard
deviation (based on the median) for the five smaller bins shown in 
Figure~\ref{6pancont}. Where necessary in 
the outer two bins, the values are given as upper limits, based on the
level of the background sky error. 
From the table and the figure, 
the integrated NMSFRs of Virgo Cluster spirals are progressively
lower in progressively outer radial bins. Significant reductions appear
first for Sa and high $C30$ in the 0.3-0.5 $r_{24}$ radial bin and
become significant
for Sb and smaller $C30$ galaxies in the 0.5-0.7$r_{24}$ bin. In the outermost
bin, reductions for all Hubble type and $C30$ bins are $>$ 3. Note
especially that Virgo galaxies classified as Sa tend to have no
massive star formation beyond 0.7$r_{24}$, and that the lower concentration
Virgo galaxies classified as Sa typically have no star formation in 
the outer half of the disk.  This latter point is another indication that
many of the Sa classifications assigned to Virgo Cluster spirals were
based on reductions in star formation rates rather than large B/D. 

The results for the outer disk are in contrast to the inner disk. Within
0.3$r_{24}$,  
95\% of Virgo Cluster spirals have NMSFRs 
similar to or enhanced compared to   
isolated galaxies of similar type or $C30$ (Figure~\ref{6pan}). 
In Table 2, the two inner bins
show isolated-to-Virgo ratios which are less than or about 1 within the
scatter for almost all type
and $C30$ bins. Note in particular the 0.38-0.50 $C30$ inner bins, where the
median NMSFR in Virgo galaxies is enhanced by a factor up to 2.5. 

In the preceding analysis, we have compared statistically only galaxies
which have $M_{R24}$ $\leq $ -19.5. In this paragraph, we summarize results
for the galaxies with lower luminosities. In Figure~\ref{maghaflux},
the total (a,b) and inner 30\% (c,d) NMSFRs of sample galaxies are
presented as a function of $M_{R24}$. We find that the behavior of
most lower luminosity spirals is similar to that of the more massive
sample galaxies: global NMSFRs are reduced, but the inner NMSFRs
are similar or mildly enhanced compared to the isolated sample.
However, the lower luminosity sample also contains galaxies for which
the global and inner rates have been significantly enhanced with
respect to this and other isolated/field samples.
The three galaxies with the highest total NMSFRs are all lower luminosity
spirals. Their NMSFRs are enhanced by a factor of 3 in the median
above isolated median rates, and they have
corresponding equivalent widths of 70-100 \AA, similar to the
starbursts M82 and NGC 1569 (Kennicutt \& Kent 1983).
The inner disk NMSFRs of lower luminosity spirals are enhanced 
even more significantly with respect to our sample of isolated galaxies: 
by a factor up to 5 for
the three galaxies with the highest global NMSFRs 
and by factors of 1.2-2.0 for several additional
galaxies. In PIV, we discuss evidence that the star formation 
rates of these galaxies have been enhanced by environmental effects.

Thus many galaxies which have
reduced global and outer disk star formation have normal or enhanced
inner disk star formation rates. These results show that the primary
mechanism of reduction in star formation for intermediate-low $C30$ galaxies
is the truncation of the outer disk rather than significant reductions 
across the disk.

\subsection{H$\alpha$ Concentration}
\label{haconc}
The tendency for Virgo Cluster galaxies 
to have truncated star-forming disks and/or enhanced inner star
formation rates means that the concentrations of 
H$\alpha$ light for these galaxies will be systematically different
than those of isolated galaxies.
We define a quantitative measure of the H$\alpha$ concentration 
as the ratio of the
flux in H$\alpha$ within 0.3$r_{24}$ to the total H$\alpha$ flux. 
The value will be higher for galaxies with truncated
star forming disks and/or enhanced inner star formation rates. In
contrast, we would not find a high value for a galaxy with 
anemic star formation, in which
the H$\alpha$ is distributed at low surface brightness over much of the disk.

\begin{figure}[t]
\includegraphics[scale=0.5]{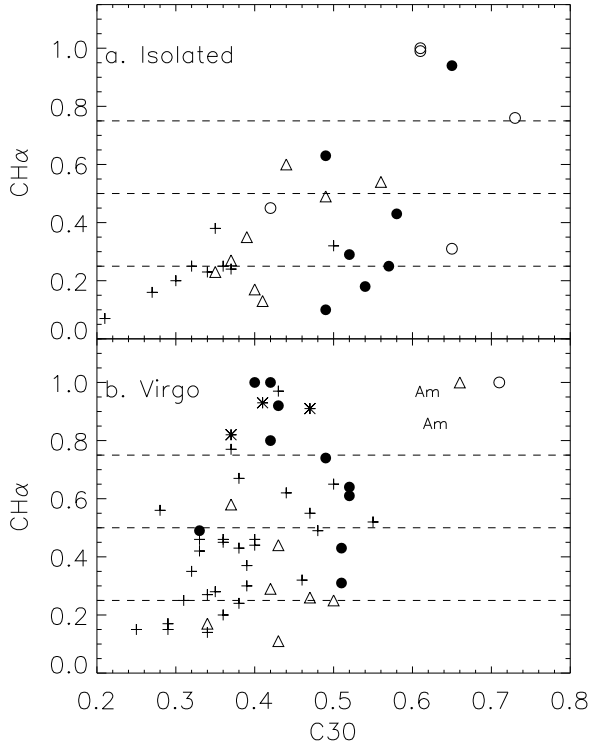}
\caption{Central R light concentration versus the 
H$\alpha$ concentration for the isolated (a) and Virgo (b) samples.
The dashed lines indicate the boundaries of
galaxies with all, half, and one-quarter of their emission within 0.3 
$r_{24}$. Symbols
indicate Hubble type, with coding the same as in Figure~\ref{6pan}.
Virgo Cluster galaxies tend to have a larger percentage of their
star formation within 0.3$r_{24}$. About 25\% of Virgo Cluster galaxies
have more than three-quarters of their star formation within 0.3$r_{24}$
compared to only 4\% of isolated galaxies. Note that no isolated low
$C30$ ($C30 \le$ 0.5) galaxy has more than 65\% of its star formation within 
0.3$r_{24}$, but that about 25\% of low $C30$ Virgo Cluster galaxies do.
This is mostly due to the truncated outer star forming disks in Virgo
Cluster galaxies. \label{hacc30}}
\end{figure}

The H$\alpha$ concentration is plotted as a function of the
R light central concentration in Figure~\ref{hacc30}. The dashed
lines indicate the boundaries
for galaxies which contain three-quarters, half,
and one-quarter of their H$\alpha$ emission within 0.3$r_{24}$.
Examination of the plot shows 
isolated galaxies typically have much of their star
formation \it outside \rm 0.3$r_{24}$, 
while Virgo galaxies more typically have most of
their star formation \it within \rm 
0.3$r_{24}$. For example, only 4\% of isolated
spirals compared to 25\% of Virgo Cluster spirals have at least
three-quarters of their H$\alpha$ emission within 0.3$r_{24}$. 
This effect is largely due to the truncated star forming disks in Virgo
Cluster galaxies.
This is not due to a morphological effect,
since the isolated sample contains proportionately \it more \rm Sa
than Virgo. 

\begin{figure}
\includegraphics[scale=0.5]{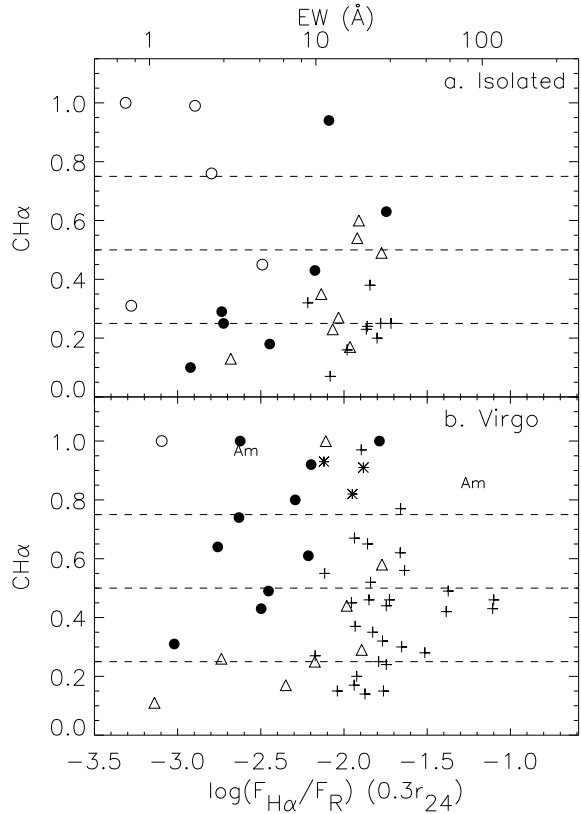}
\caption{H$\alpha$ concentration vs the inner normalized 
H$\alpha$ flux for the isolated (a) and Virgo (b) samples. The dashed lines 
indicate the boundaries of galaxies with all, half, and one-quarter of 
their emission within 0.3  $r_{24}$. Symbols
indicate Hubble type, with coding the same as in Figure~\ref{6pan}.
The upper x-axis provides the equivalent width scale.
The inner star formation rates of Virgo Cluster
spirals are similar to those of isolated spirals, but Virgo Cluster
galaxies tend to have higher CH$\alpha$. Thus Virgo spirals tend to 
have normal-enhanced inner star formation combined with truncated outer
disks.\label{nhaf3hac}}
\end{figure}

The H$\alpha$ concentration does not reveal the relative level of star
formation of a galaxy. In Figure~\ref{hacc30}, for example, 
S0 galaxies with reduced
star formation have high H$\alpha$ concentration values and therefore
fall in similar locations as galaxies with active inner star formation
and truncated disks. By plotting the H$\alpha$ concentration versus the 
level of inner normalized H$\alpha$ flux, as in Figure~\ref{nhaf3hac},
it is apparent that 
a large proportion of Virgo Cluster galaxies have inner star formation
rates similar to those of isolated Sb-Sc galaxies, but have much higher
H$\alpha$ concentration.
Thus the Virgo galaxies tend to have inner disks with normal to enhanced 
star formation combined with truncated outer disks. 
In PIV, we show that the H$\alpha$ concentration can be used with
the global NMSFR as an
indicator of the star formation morphology of a spiral galaxy.

\section{The Star Formation Rates of HI Deficient Galaxies}

Virgo Cluster spiral galaxies have long been known to have less HI gas
than field counterparts (Chamaraux et al 1980; Giovanelli \& Haynes 1983).
Previous studies (e.g., Chamaraux, Balkowski, \& Fontanelli 1986; 
Solanes et al. 2001; see also 
Kenney 1990) concluded that the reduction in HI gas, quantified with
an HI deficiency parameter (Giovanelli \& Haynes 1983; Solanes et al. 1996), 
has been
more severe for early-type spiral galaxies. Based on this result,
Chamaraux et al. (1986) and van Driel (1987) proposed that HI gas in
early-type galaxies is more likely to be located in the outer disk, where
it is easier to strip. Dressler (1986) suggested that HI deficient galaxies,
which tend to have early-type classifications,
are more likely to have radial orbits than HI rich galaxies, 
based on analysis of the velocity dispersions
of HI rich and HI deficient galaxies in 9 clusters. 

\begin{figure}
\includegraphics[scale=0.5]{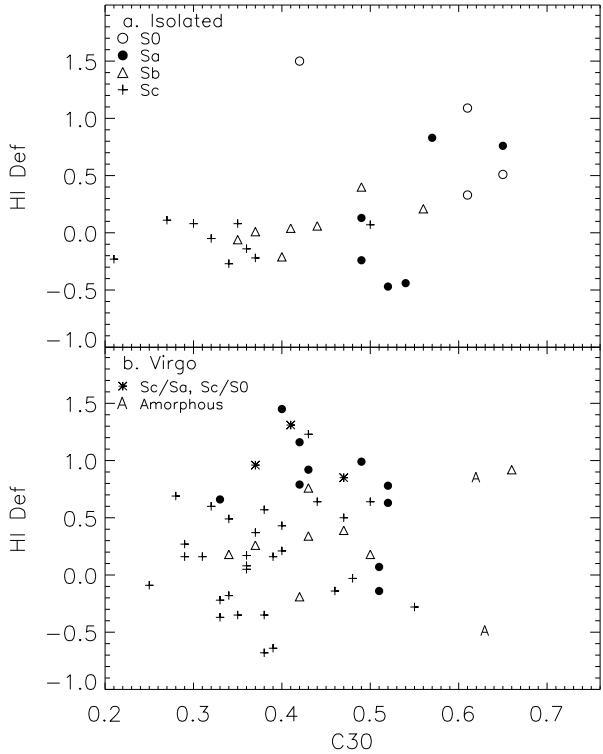}
\caption{HI deficiency as a function of $C30$ for the isolated (top) and
Virgo (bottom) galaxies. The HI deficiency values were calculated
following the prescription set out in Solanes et al. (1996). On this
scale, a galaxy with normal HI content has an HI deficiency value of
0, while a galaxy which is depleted by a factor of 10 has an HI
deficiency value of 1.0. The symbols indicate Hubble type. Note that there
is no correlation between HI deficiency and $C30$ or HI deficiency and
Hubble type for the isolated sample. However, there is a clear tendency
for HI deficient Virgo Cluster galaxies to be classified as earlier
types, despite their central concentrations.
\label{c30hidef}}
\end{figure}

In light of the presence of a number of galaxies in the Virgo Cluster
with misleading classifications, it is worth reexamining the 
relationship between HI deficiency and morphology.
HI deficiencies were calculated following Solanes et al. (1996; see
also Giovanelli \& Haynes 1983). The values presented here for the Virgo
Cluster have been updated from those presented in PI and are tabulated in PIV.
HI fluxes were extracted from the Cornell University Extragalactic 
Group private digital HI archive, and are corrected for effects such as beam
dilution (Springob, Haynes, \& Giovanelli, in prep; see also Haynes \&
Giovanelli 1984). Optical diameters were obtained from the Arecibo 
General Catalog, a private database maintained at Cornell University
by Martha Haynes and Riccardo Giovanelli. Distances are based on the
heliocentric velocity and the multiattractor model for the velocity
fields within the Local Supercluster (Tonry et al. 2000; Masters, K. L.,
private communication), with an assumed Hubble constant of 70 km s$^{-1}$ 
Mpc$^{-1}$.

In Figure~\ref{c30hidef}, we plot HI deficiency versus $C30$, marking the
points according to Hubble type, for the isolated and Virgo Cluster
samples. The Virgo Cluster sample has many HI deficient galaxies: 
there is
a large population of galaxies with HI depleted by at least a factor
of 5. It also has several galaxies with HI enhanced by a factor of
a few. In the isolated sample, there is almost no correlation between
HI deficiency and either $C30$ or Hubble type (except that most of the galaxies
with depleted HI are  highly concentrated galaxies classified
as S0 or Sa). In the Virgo sample, there is also no correlation
between HI deficiency and $C30$.
However, there is a stronger correlation
between Hubble type and HI deficiency, in that many of the galaxies
with HI deficiency greater than $\sim$ 0.8 are classified as Sa, regardless
of their central light concentration. 
It is thus clear from this plot that \it the
correlation between HI deficiency and central concentration is
much weaker than the correlation between HI deficiency and Hubble
type. \rm

\begin{figure}
\includegraphics[scale=0.5]{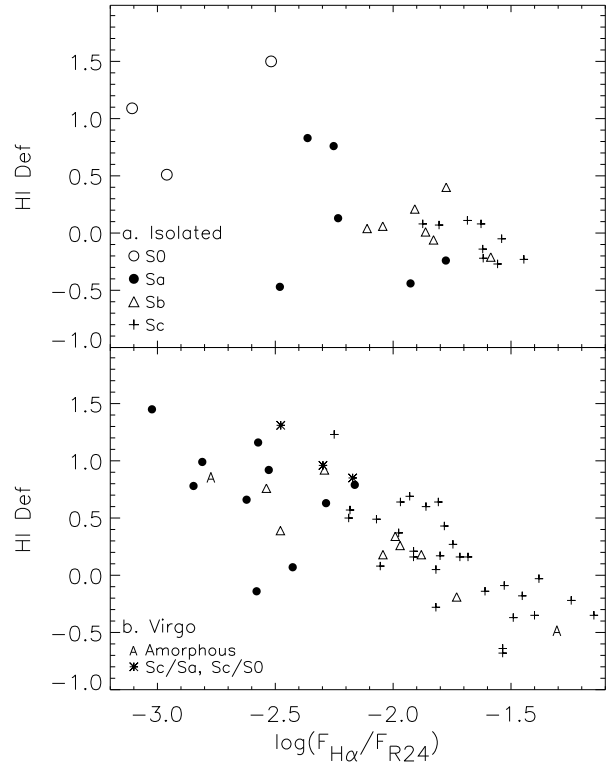}
\caption{HI deficiency as a function of total normalized star formation 
rates 
for the isolated (top) and Virgo (bottom) galaxies. There is a correlation 
between total normalized H$\alpha$ fluxes and HI deficiencies in the Virgo 
sample, markedly stronger than that between $C30$ and HI deficiency.
This suggests that gas-poor galaxies with reduced star formation 
tend to be classified as early-type spirals.\label{hafluxhidef}}
\end{figure}

The correlation between
Hubble type and HI deficiency is really a reflection of a correlation
between star formation rates and HI deficiency in the Virgo Cluster. 
In Figure~\ref{hafluxhidef}, the normalized H$\alpha$ fluxes are plotted
versus the HI deficiency parameter, with symbols marked according
to Hubble type. In the Virgo Cluster, there is a strong correlation 
between total normalized H$\alpha$ flux and HI deficiency (see also
Gavazzi et al. 2002), with the assignment of Hubble type tending in the 
same direction. 

The stronger correlation between Hubble type and HI deficiency is
one of the biases caused by forcing cluster galaxies into a 
one-dimensional classification system. 
Figures~\ref{c30hidef} and ~\ref{hafluxhidef}
together show that galaxies of any bulge-to-disk ratio
can be strongly stripped of HI, suffer reduced star formation,
and be classified as early type spirals.
Therefore, the correlation between HI deficiency and Hubble
type is no longer surprising, rather it is 
predominantly due to strongly stripped galaxies being classified as Sa's,
rather than galaxies with large bulge-to-disk ratios being
preferentially stripped. We therefore suggest that
the reason for the previously observed systematic difference
in spatial distributions and kinematics between early- and late-type
Virgo spirals is at least partially due to the fact 
that stripped spirals, which are preferentially on radial orbits
(Solanes et al. 2001),
tend to be misleadingly \it classified \rm as early-types.

\section{The Effects of Extinction on H$\alpha$ Results}

H$\alpha$ will underestimate the true massive star formation rate
due to extinction by dust. 
Since we have done a comparative study
of Virgo Cluster and isolated galaxies, our results will be affected only if
there is a systematic difference in extinction between the two samples.

There could be a systemic difference, since stripped galaxies likely have less
dust as well as less gas.
There are likely radial gradients in extinction, with more in the centers
than the outer disks (Giovanelli et al. 1994, 1995).
Stripping and truncation selectively removes the low extinction part 
of the disks, i.e. the  outer gas and dust disks.
So the true difference in star formation rates
between the stripped and unstripped galaxies
may be less than is indicated in H$\alpha$.
Any difference should be less in the comparisons within distinct radial bins.
The fact that we see large differences between the Virgo and isolated
galaxies in the outer radial bins shows clearly that many outer Virgo
disks are extremely deficient in H$\alpha$.
However, the outer disk contains a smaller fraction of the global star
formation rate than is
indicated by uncorrected H$\alpha$, and the true global reduction in star
formation rate may be less than is indicated by the H$\alpha$ results.

Tracers of SF which measure dust emission, such as FIR, will be subject to
the opposite problem of overemphasizing the inner disk.
An improved treatment should use information from both optical and FIR
emission to estimate the true radial profile of star formation.
\section{Discussion}

\subsection{General Comments on the Morphology of Virgo Cluster spirals}
Classification is inherently subjective and, 
as we and others have shown in the case of Virgo Cluster galaxies,
can be dangerous and misleading.
Objectives measures are generally preferable, but
classification can have value, 
so here we offer some comments on an improved classification
for large disk galaxies.

The standard Hubble classification scheme, as employed by
Sandage and deVaucouleurs, uses both bulge-to-disk ratio
and disk structure (closely related to disk star formation rate).
This system cannot work for cluster galaxies because
in Virgo the bulge-to-disk ratio is poorly correlated with
the disk star formation rate.
As pointed out by  van den Bergh (1976), the Sa type has become 
a repository for galaxies with very different physical characteristics.

The van den Bergh (1976) RDDO system offers an important step in
a more correct approach,
by completely separating the spatial distribution of the old stars
(i.e. bulge-to-disk ratio) from the 
the spatial distribution of the new stars
(as measured, for example, by H$\alpha$).
The three proposed evolutionary stages (normal spiral, anemic, and lenticular) 
are still not sufficient to describe the morphology
of approximately half of the Virgo spirals in this sample, which have spatially
truncated star formation
profiles: a normal-enhanced inner disk star formation combined with weak-no
outer disk star formation.
Whereas many isolated and some Virgo galaxies might follow an evolutionary
progression from normal spiral, through anemic, and then to lenticular,
many cluster galaxies will not.

In broadband images, truncation of the star-forming disks results in a
smooth outer disk combined with a clumpy inner disk. This effect was noticed
by van den Bergh et al. (1990), who called galaxies with this morphology
"Virgo types". While this morphology reflects the most frequently occurring type of environmental alteration in the Virgo
Cluster, there are other peculiar morphologies in the Virgo
Cluster, and so we have chosen not to use the term `Virgo-type'.
We propose that an additional parallel branch could be added to the van
den Bergh scheme, describing the truncated star formation morphology: 
STa - STb -STc, where 
the a-b-c division is based on the central light concentration or B/D.
However, we reemphasize that attempts to characterize galaxy morphologies
should be founded on objective measurements rather than subjective
visual classification whenever possible. Studies such as Conselice (2003) 
offer promise for better characterizing galaxy morphologies in terms of a few 
relatively simple global 
parameters such as central concentration and asymmetry, 
which Conselice shows are correlated to a galaxy's evolutionary state. 
The truncation phenomena, which seems most prevalent in cluster galaxies,
should be included in any such objective parameter schemes.

\subsection{Comparison to Previous Work on the Star Formation
Properties of Cluster Galaxies}

Most previous studies of cluster galaxies have been limited to measurements of 
global star formation. Comparisons between cluster and field samples produce
varying results, as documented in the introduction. 
It is clear that the star formation activity in the Virgo Cluster spans a
large range. We find, in agreement with a number of previous multiwavelength
studies of Virgo (Kennicutt 1983a; Kodaira et al. 1990; Gavazzi et al. 2002),
that the global star formation activity as traced by H$\alpha$ 
has been reduced in the median. The amount of reduction
may well be greater, since our study includes 
mostly spirals with at least some ongoing star formation activity. The 
selection of sample galaxies using the Hubble type may also lead to an
underestimate in the reduction in star formation, 
since small-bulge galaxies which 
have ceased star formation are likely classified as S0's. 

The result that several Virgo Cluster spirals have enhanced star
formation is relatively new for studies of the Virgo Cluster.
This may be partly due to previous comparison samples.
The Kennicutt \& Kent (1983) study of H$\alpha$ emission in Virgo cluster
galaxies, for example, used a
comparison sample of field galaxies, which included systems in groups,
and in particular some starbursts (e.g., M82).
Thus their study was not especially sensitive to 
rates of star formation enhanced above those found in isolated galaxies,
which we are better able to quantify. 
We note that
most of the Virgo Cluster galaxies
with enhanced star formation have absolute B magnitudes between -18 and -19.
This result is probably related to the observation that `k+a' 
and emission-line galaxies in the Coma Cluster 
tend to be less luminous than `k+a' counterparts at larger redshifts (Poggianti
et al. 2004).

The Moss \& Whittle (2000) study finds a high percentage
of starbursting late-type spirals in 8 nearby Abell clusters.
However, it is important to note that 
their survey was not sensitive to reductions
in star formation activity because the survey has a rather high lower limit of 
$EW(H\alpha) \simeq 20 \AA$ on the strength of H$\alpha$.
(Most of the galaxies in our sample have $EW(H\alpha$) below 20 $\AA$.)
This means Moss \& Whittle cannot detect differences between normal and
reduced amounts of star formation, and they cannot determine
whether the overall star formation rate in their clusters
is higher or lower than non-cluster samples.

There are 
relatively few previous results on the spatial distributions of star 
formation for galaxies in clusters. Our work shows 
that most galaxies with reduced star formation in the Virgo Cluster have 
truncated star-forming disks, but normal-to-enhanced inner rates. 
Rotation curve studies (Rubin et al. 1999; Dale et al. 2001) also show
the truncated H$\alpha$ distributions of cluster galaxies. Dale et al. (2001)
find on average more severely truncated galaxies closer to the cluster core
(see also PIV).

Other studies have found evidence of centrally concentrated star formation
morphologies in a number of cluster galaxies, which may be combined
with a severe truncation of the star-forming disk.
Caldwell et al. (1996) identified a population of starburst and poststarburst
spirals in Coma, most of which have centrally concentrated star formation, 
typically within 1-2 kpc (Caldwell et al. 
1999). Similar morphology galaxies are found in Pegasus I (Rose et al. 2001). 
These poststarburst galaxies were initially classified as E or
S0, and yet all are spirals and about half have small-intermediate B/D.
In a study of the S0's in Coma, Poggianti et al. (2001) found that 40\% 
show spectroscopic evidence for central star formation within the last 5 Gyr. 
Finally, Moss \& Whittle (2000) find evidence of circumnuclear starbursts 
in a population of cluster galaxies.

Thus, 
this study and those cited above show that truncated star-forming disks with
active inner star formation are prevalent in clusters, but less common
in the field, suggesting that they are produced by
environmental processing in clusters.

\subsection{Environmental Effects in the Virgo Cluster}
Studies of the spatial characteristics of star formation 
in the Virgo and other clusters are especially important because of the 
clues they provide in investigating the dominant environmental effects
on galaxy evolution in clusters. In this section, we summarize the 
implications of this survey on the understanding of environmental
effects in the Virgo Cluster. A more extensive discussion is given
in PIV, where we investigate possible
relationships between observed H$\alpha$ morphologies and environmental
effects.

The systematic reduction in the star formation rate in 
the outer disks of many Virgo spirals revealed by this study 
is in all likelihood due to the systematic
removal of gas from the outer disk by the Virgo environment. 
Star formation ceases once the gas falls below a threshold determined by the
local dynamics (Kennicutt 1989). 
The sharp spatial truncation of the H$\alpha$ disks in such a high
percentage of Virgo Cluster galaxies is most naturally explained by
gas removal due to ICM-ISM interactions (e.g. Gunn \& Gott; Nulsen 1982;
Vollmer et al. 2001).
In the case of starvation (Larson et al. 1980; Balogh, Navarro, \& Morris
2000), i.e., the stripping of a loosely-bound outer gas 
reservoir (by the ICM and/or cluster potential field), one might expect to
see a gradual reduction in star formation across the disk, resulting
in a significant fraction of anemic spirals, which we show to
be rare in the Virgo Cluster (see also PIV). 
While tidal interactions may be capable
of eventually producing spirals with truncated gas disks (see for example,
Kenney et al. 1995) and may be important in the evolution of many Virgo
spirals (PIV), it is hard to explain how tidal interactions alone could account
for such a large population
of galaxies with spatially truncated gas disks. In addition, we show in 
PIV that many of the galaxies with truncated H$\alpha$ disks have regular 
stellar isophotes. 

If ICM-ISM stripping is an important effect in the Virgo Cluster, we would 
expect to see gas- and star-formation-poor galaxies with a continuum of
bulge sizes. Our observations show that this is the case; in particular,
we find significant numbers of small to intermediate bulge-to-disk ratio
galaxies with truncated star forming disks.

The normal to enhanced star formation rates in
the inner galaxy disks of Virgo spirals imply that ICM-ISM interactions have 
not had a strong
effect on the inner regions of most Virgo (star-forming) galaxies. 
The inner star formation rates of even the severely HI deficient spirals
are within the bounds of isolated spirals. This is consistent with the
observation that the inner disks of galaxies are H$_2$-dominated,
and that CO luminosities are close to normal for Virgo galaxies compared
to field galaxies
(Kenney \& Young 1989). Apparently the total gas surface density in inner
disks is still high enough to allow star formation.  While
it has been suggested that enhancements in star formation may be
caused by ICM-ISM interactions (e.g., Dressler \& Gunn 1983; Gavazzi et al.
2001), 
it appears that our sample galaxies with enhanced global and inner disk
star formation have been more strongly affected by processes 
other than ICM-ISM stripping (PIV).

Thus the results of this survey support the conclusion that
ICM-ISM interactions play a significant role in the evolution of many
Virgo Cluster spirals, by preferentially stripping gas from the outer disk. 
More detailed study of the H$\alpha$ morphologies
presented in PIV supports this conclusion and also indicates that tidal 
encounters and mergers are also important in explaining the
optical and kinematical peculiarities (e.g.,
Rubin et al. 1999) of a number of Virgo Cluster spirals.

\section{Conclusions}

The results of this survey of star formation rates and morphologies
of Virgo Cluster spirals compared to isolated spirals include the following:

1. H$\alpha$-based estimates of the total massive star formation 
rates of Virgo Cluster galaxies 
span a range from strongly reduced (up to 10 times) to enhanced 
(up to 2.5 times) compared to the 
isolated sample. In the median, Virgo total star formation 
rates are reduced by factors up to 2.5 for different Hubble
types and concentrations.

2. For most Virgo Cluster galaxies with reduced total star formation, it is 
truncation rather than anemia (low H$\alpha$
surface brightness across the disk) which causes
the reduced total star formation rates. 
Median inner rates are similar or enhanced up to a factor of 1.7,
while outer star formation rates are reduced in the median by factors
of 1.5 - 9. In the outermost parts of the optical disks, star formation of
all types and concentrations of galaxies are reduced by factors greater
than 3 times in the median. 
Reductions in individual galaxies range to factors greater
than 100. No galaxies have star formation rates in the outer half of
the disk enhanced above isolated rates. Thus, most
galaxies with reduced total star formation have inner star formation
rates which are similar to or enhanced with respect 
to isolated galaxies of similar central light concentration or Hubble type. 
Virgo Cluster galaxies have more concentrated star forming disks than
isolated counterparts, largely due to truncation. 

3. The larger mean HI deficiency of Virgo cluster Sa's 
as compared to cluster Sb and Sc's (Solanes et al 2001)
is  predominantly due to strongly stripped galaxies being classified as Sa's,
rather than galaxies with large bulge-to-disk ratios being
preferentially stripped.
One reason for the previously observed systematic difference
in spatial distributions and kinematics between early- and late-type
Virgo spirals (Dressler 1986) is 
that stripped spirals, which are preferentially on radial orbits
(Solanes et al 2001),
tend to be misleadingly \it classified \rm as early-types. 
Likewise, the excess of Sa galaxies in the Virgo cluster,
and perhaps by extension other clusters,
is in large part due to strongly stripped galaxies being classified as Sa's,
and is not simply an excess of spiral galaxies with large bulge-to-disk ratios.

The funding for the research on the Virgo cluster
and isolated spiral galaxies was provided by NSF grants AST-9322779
and AST-0071251.
Martha Haynes, Christopher Springob, Karen Masters, and the Cornell 
Extragalactic Group are gratefully acknowledged for their aid in 
derivation of updated HI deficiencies.
We thank Judy Young, Vera Rubin, Yasuhiro Hashimoto, and Shardha
Jogee for valuable discussions, and our referee, Alessandro Boselli, 
for helpful comments which improved this paper.
This research has made use of the NASA/IPAC Extragalactic Database (NED)
which is operated by the Jet Propulsion Laboratory, California Institute
of Technology, under contract with the National Aeronautics and Space
Administration.


\begin{references}
\reference{} Abraham, R. G., Valdes, F., Yee, H. K. C., \& van den Bergh, S. 1994, ApJ, 432, 75
\reference{} Abraham, R. G., Smecker-Hane, T. A., Hutchings, J. B., Carlberg, R. G., Yee, H. K. C., Ellingson, E., Morris, S., Oke, J. B., \& Rigler, M. 1996,
ApJ, 471, 694
\reference{} Balogh, M. L., Navarro, J. F., \& Morris, S. L. 2000, ApJ,
540, 113
\reference{} Balogh, M. L., Schade, D., Morris, S., Yee, H. K. C., Carlberg, 
R. G., \& Ellingson, E. 1998, ApJ, 504, 75
\reference{} Bennett, S.M. \& Moss, C. 1998, A\&A, 132, 55
\reference{} Bicay, M. D. \& Giovanelli, R. 1987, ApJ, 321, 645
\reference{} Binggeli, B., Sandage, A., \& Tammann, G. A. 1985, AJ, 90, 1681 (BST)
\reference{} Boselli, A., Gavazzi, G., Donas, J., \& Scodeggio, M. 2001, AJ, 121, 753
\reference{} Bothun, G. D. 1982, PASP, 94, 774
\reference{} Butcher, H. \& Oemler, A., Jr.  1978, ApJ, 219, 18
\reference{} Caldwell, N., Rose, J. A., Franx, M., \& Leonardi, A. 1996, AJ, 111, 78
\reference{} Caldwell, N., Rose, J. A., \& Dendy, K. 1999, AJ, 117, 140
\reference{} Cayatte, V., van Gorkom, J. H., Balkowski, C., \& Kotanyi, C. 1990, AJ, 100, 604
\reference{} Chamaraux, P., Balkowski, C., \& G\'erard, E. 1980, A\&A, 83, 38
\reference{} Chamaraux, P., Balkowski, C., \& Fontanelli, P. 1986, A\&A, 165, 15
\reference{} Conselice, C. J. 2003, ApJS, 147, 1
\reference{} Dale, D. A., Giovanelli, R., Haynes, M. P., Hardy, E., Campusano, L. E. 2001, ApJ, 549, 215  
\reference{} deVaucouleurs, G., deVaucouleurs, A., Corwin, H. G., Buta, R. J., Paturel, G., \& Fouqu\'{e}, P. 1991, Third Reference Catalog of Bright Galaxies (New York: Springer-Verlag)
\reference{} Donas, J, Buat, V., Milliard, B., \& Laget, M. 1990, A\&A, 235, 60
\reference{} Dressler, A. 1980, ApJ, 236, 351
\reference{} Dressler, A. 1986, ApJ, 301, 35
\reference{} Dressler, A.\& Gunn, J. E. 1983, ApJ, 270, 7
\reference{} Dressler, A., Oemler, A., Couch, W. J., Smail, I., Ellis, R. S., Barger, A., Butcher, H., Poggianti, B. M., \& Sharples, R. M. 1997, ApJ, 490, 577
\reference{} Gavazzi, G., Boselli, A., \& Kennicutt, R. 1991, AJ, 101, 1207
\reference{} Gavazzi, G., Catinella, B., Carrasco, L., Boselli, A.,
Contursi, A. 1998, AJ, 115, 1745
\reference{} Gavazzi, G., Boselli, A., Mayer, L., Iglesias-Paramo, J., Vilchez,  J.M. \& Carrasco, L. 2001, ApJ, 563, L23
\reference{} Gavazzi, G., Boselli, A., Pedotti, P., Gallazzi, A. \& Carrasco, L. 2002, A\&A, 396, 449
\reference{} Giovanelli, R. \& Haynes, M. P. 1983, AJ, 88, 881
\reference{} Giovanelli, R., Haynes, M. P., Salzer, J. J., Wegner, G., Da Costa, L. N., \& Freudling, W. 1994, AJ, 107, 2036
\reference{} Giovanelli, R., Haynes, M. P., Salzer, J. J., Wegner, G., Da Costa, L. N., \& Freudling, W. 1995, AJ, 110, 1059
\reference{} Gourgoulhon, E., Chamaraux, P., \& Fouqu\'{e}, P. 1992, A\&A, 255,69
\reference{} Graham, A. W. 2001, AJ, 121, 820
\reference{} Graham, A. W., Trujillo, I., \& Caon, N. 2001, AJ, 122, 1707
\reference{} Gunn, J. E. \& Gott, J. R. 1972, ApJ, 176, 1
\reference{} Hameed, S. \& Devereux, N. 1999, AJ, 118, 730
\reference{} Hashimoto, Y., Oemler, A., Lin, H., \& Tucker, D. L. 1998, ApJ, 499. 589
\reference{} Haynes, M. P. \& Giovanelli, R. 1984, AJ, 89, 758
\reference{} Hubble, E. \& Humason, M. L. 1931, ApJ, 74, 43
\reference{} Kenney, J. D. P. 1990, in The Interstellar Medium in Galaxies, ed. H. A. Thronson \& Shull, J.M. (Dordrecht: Kluwer), 151
\reference{} Kenney, J.D.P. \& Young, J. 1989, ApJ, 344,171
\reference{} Kennicutt, R. C. 1983a, AJ, 88, 483
\reference{} Kennicutt, R. C. 1983b, ApJ, 272, 54
\reference{} Kennicutt, R. C. 1989, ApJ, 344, 685
\reference{} Kennicutt, R. C. 1998, ARA\&A, 36, 189
\reference{} Kennicutt, R. C., Bothun, G. D., \& Schommer, R. A. 1984, AJ, 89, 179
\reference{} Kennicutt, R. C. \& Kent, S. M. 1983, AJ, 88, 1094
\reference{} Kent, S. M. 1985, ApJS, 59, 115
\reference{} Kodaira, K., Watanabe, T., Onaka, T, \& Tanaka, W. 1990, ApJ, 363, 422
\reference{} Koopmann, R.A. \& Kenney, J.D.P. 1998, ApJ, 497, L75
\reference{} Koopmann, R. A.,  Kenney, J. D. P., Young, J. 2001, ApJS, 135, 125 (PI)
\reference{} Koopmann, R.A. \& Kenney, J.D.P. 2004, submitted (PIV)
\reference{} Larson, R. B., Tinsley, B. M., \& Caldwell, C. N. 1980,
ApJ, 237, 692
\reference{} Mihos, J. C. 2004, in ``Clusters of Galaxies: Probes of Cosmological Structure and Galaxy Evolution'', ed. J. S. Mulchaey, A. Dressler, \& A. Oemler (Cambridge: Cambridge Univ. Press), 278
\reference{} Miller, R. H. 1988, Comm.Astrop. 13, 1
\reference{} Moore, B., Lake, G., \& Katz, N. 1998, ApJ, 495, 139
\reference{} Moss, C. \& Whittle, M. 1993, ApJ, 407, L17
\reference{} Moss, C., \& Whittle, M. 2000, MNRAS, 317, 667
\reference{} Nulsen, P. E. J. 1982, MNRAS, 198, 1007
\reference{} Oemler, A. 1974, ApJ, 194, 1
\reference{} Oemler, A., Jr. 1992 in Clusters and Superclusters of Galaxies, ed. A.C. Fabian (Dordrecht: Kluwer), 29
\reference{} Poggianti, et al.  2001, ApJ, 563, 118
\reference{} Poggianti, B. M., Bridges, T. J., Komiyamia, Y., Yagi, M., Carter, D., Mobasher, B., Okamura, S., \& Kashikawa, N. 2004, ApJ, 601, 197
\reference{} Postman, M.  \& Geller, M. J. 1984, ApJ, 281, 95
\reference{} Rose, J. A., Gaba, A. E., Caldwell, N., \& Chaboyer, B. 2001, AJ 121, 793
\reference{} Rubin, V.C., Waterman, A.H., \& Kenney, J.D.P. 1999, AJ, 118,236
\reference{} Sandage, A., \& Bedke, J. 1994, The Carnegie Atlas of Galaxies (Washington: Carnegie)
\reference{} Sandage, A., \& Tammann, G. A. 1987, A Revised Shapley-Ames Catalog of Bright Galaxies (Washington: Carnegie)
\reference{} Scodeggio, M., Gavazzi, G., Franzetti, P., Boselli, A., Zibetti, S., \& Pierini, D. 2002, A\&A, 384, 812
\reference{} Solanes, J. M., Giovanelli, R., \& Haynes, M. P. 1996, ApJ, 461, 609
\reference{} Solanes, J. M., Manrique, A., Garcia-Gomez, C., Gonzalez-Casado,
 G., Giovanelli, R., \& Haynes, M. P. 2001, ApJ, 548, 97
\reference{} Struck, C. 1999, Physics Reports, 321, 1
\reference{} Tonry, J. L., Blakeslee, J. P., Ajhar, E. A., \& Dressler, A. 2000, ApJ, 530, 625
\reference{} Tully, R. B. \& Shaya, E. J. 1984, ApJ, 281, 31
\reference{} Tully, R. B. 1987, Nearby Galaxies Catalog (Cambridge: Cambridge University Press) 
\reference{} van den Bergh, S. 1976, ApJ, 206, 883
\reference{} van den Bergh, S., Pierce, M.J., \& Tully, R. B. 1990, ApJ, 359, 4
\reference{} van Driel, W., 1987, PhD thesis, Groningen
\reference{} van Gorkom, J. H., in ``Clusters of Galaxies: Probes of Cosmological Structure and Galaxy Evolution'', ed. J. S. Mulchaey, A. Dressler, \& A. Oemler (Cambridge: Cambridge Univ. Press), 306
\reference{} Vollmer, B., Cayatte, V., Balkowski, C., \& Duschl, W. J.
  2001, ApJ, 561, 708
\reference{} Warmels, R. H. 1988, A\&AS, 72, 19
\reference{} Young, J.S., Allen, L., Kenney, J.D.P., Lesser, A., \& Rownd, B. 1996, AJ, 112, 1903
\end{references}
\end{document}